\documentclass[prd,twocolumn,showpacs,preprintnumbers,superscriptaddress,floatfix,nofootinbib]{revtex4-1}
\usepackage{graphicx,,booktabs}
\usepackage{epstopdf}
\graphicspath{{figures/}}
\usepackage{amssymb}
\usepackage{amsmath,txfonts}
\usepackage{graphicx}
\usepackage{dcolumn}
\usepackage{color}
\usepackage{bm}
\usepackage[colorlinks,
            citecolor=blue,
            anchorcolor=green,
            menucolor=orange,
            linkcolor=red,
            filecolor=red,
            runcolor=pink,
            urlcolor=blue,
            frenchlinks=red]{hyperref}
\usepackage{multirow}
\usepackage[capitalize]{cleveref}
\usepackage{slashed}
\makeatletter

\newcommand{\Rmnum}[1]{\expandafter\@slowromancap\romannumeral #1@}

\makeatother

\begin{document}

\title{\boldmath Producing $\Lambda(1405)$ and $\Lambda(1520)$ in $\pi^-p$ reaction to explore their inner structures}

\author{Yuan Gao}
	\affiliation{Department of Physics, Lanzhou University of Technology,
		Lanzhou 730050, China}
        
\author{Xiao-Yun Wang}
	\email{xywang@lut.edu.cn}
	\affiliation{Department of Physics, Lanzhou University of Technology,
		Lanzhou 730050, China}
	\affiliation{Lanzhou Center for Theoretical Physics, Key Laboratory of Theoretical Physics of Gansu Province, Lanzhou University, Lanzhou, Gansu 730000, China}
	
	\author{Xiang Liu}
	\email{xiangliu@lzu.edu.cn}
	\affiliation{Lanzhou Center for Theoretical Physics, Key Laboratory of Theoretical Physics of Gansu Province, Lanzhou University, Lanzhou, Gansu 730000, China}
	\affiliation{School of Physical Science and Technology, Lanzhou University, Lanzhou 730000, China}
	\affiliation{Key Laboratory of Quantum Theory and Applications of MoE, Lanzhou University, Lanzhou 730000, China}
	\affiliation{MoE Frontiers Science Center for Rare Isotopes, Lanzhou University, Lanzhou 730000, China}
	\affiliation{Research Center for Hadron and CSR Physics, Lanzhou University and Institute of Modern Physics of CAS, Lanzhou 730000, China}

\begin{abstract}
In this work, the production mechanisms of the hyperon resonances $\Lambda(1405)$ and $\Lambda(1520)$ in the $\pi^- p$ scattering are investigated within an effective Lagrangian approach incorporating Regge trajectories. By including contributions from $t$-channel $K^*$ and $u$-channel $\Sigma$ exchanges, we perform global fits to the total and differential cross sections for $\pi^{-} p \rightarrow K\Lambda(1405)$ and $\pi^{-} p \rightarrow K\Lambda(1520)$. The results show good agreement with available experimental data. For the total cross section of $\Lambda(1405)$ production, the $u$-channel contribution is dominant, whereas the $t$-channel contribution plays the primary role in $\Lambda(1520)$ production. Furthermore, the differential cross sections of the two processes exhibit distinctly different shapes, reflecting their distinct underlying reaction mechanisms. An analysis based on the constituent counting rule indicates that $\Lambda(1520)$ is consistent with a conventional three-quark configuration, while $\Lambda(1405)$ shows a clear deviation, suggesting a more exotic structure. Owing to the large branching ratio of $\Lambda^* \to \pi \Sigma$, the Dalitz process $\pi^{-} p \rightarrow K \Lambda^{*} \rightarrow K \pi \Sigma$ is also calculated. Our results demonstrate that reconstructing $\Lambda^*$ via the $K\pi\Sigma$ final state is experimentally feasible. This study provides important theoretical insights into the production dynamics of these hyperon resonances, and suggests future high-precision measurements of the $t$-distribution at large momentum transfer at facilities such as AMBER, J-PARC, HIKE, and HIAF, which can further clarify their reaction mechanisms and structural properties.
\end{abstract}
\maketitle
\section{INTRODUCTION}

Studies in hadron spectroscopy serve as a crucial approach to deepening our understanding of the nonperturbative behavior of strong interactions~\cite{ParticleDataGroup:2024cfk,Wang:2023poi,WANG:2025fmh}, which remains a core unresolved issue in particle physics. Within the hadron family, hyperons represent an important subject of research~\cite{ParticleDataGroup:2024cfk,WANG:2025fmh}.

As two of the most extensively studied excited states among light-flavored hyperons, $\Lambda(1405)$ and $\Lambda(1520)$ are central objects in hyperon spectroscopy.
The former, whose mass lies below the $\bar{K}N$ threshold and cannot be readily described by simple three-quark models, has long been debated in terms of its internal structure—whether it is a meson–baryon molecular state or a multiquark mixed configuration~\cite{Alston:1961zzd, ParticleDataGroup:2024cfk, Dalitz:1967fp, Isgur:1978xj, Kaiser:1996js, Oller:2000fj, Jido:2003cb, Guo:2012vv, Meissner:2020khl, Xie:2023jve, Ikeda:2012au, Lu:2022hwm, Nemoto:2003ft, An:2010wb}.
The latter, a typical spin-orbit excited state ($J^P = 3/2^-$), is often regarded as the parity partner of the $\Lambda$ ground state; its distinctive decay channels and spectroscopic features provide crucial benchmarks for testing models of baryon structure.
Differences in their spectra, decay behaviors, and structural nature not only reflect the diversity of light-flavored hyperon excitations but also offer complementary experimental and theoretical probes into the dynamics of quark confinement and strong interactions in the non-perturbative regime of QCD.

Over the past decades, several experiments have measured the photoproduction cross sections of $\gamma p \to K^+\Lambda(1405)$ and $\gamma p \to K^+\Lambda(1520)$.
For $\gamma p \to K^+\Lambda(1405)$, high-precision differential and total cross-section data were reported by the CLAS Collaboration at Jefferson Lab in the center-of-mass energy range $W \approx 2.0$–$2.8\ \text{GeV}$~\cite{CLAS:2013rxx}.
For $\gamma p \to K^+\Lambda(1520)$, measurements were performed by Boyarski \textit{et al.} at SLAC~\cite{Boyarski:1970yc}, the LAMP2 group~\cite{Barber:1980zv}, the LEPS Collaboration at SPring-8~\cite{LEPS:2009isz, Muramatsu:2009zp}, the SAPHIR Collaboration at ELSA~\cite{Wieland:2010run}, and the CLAS Collaboration~\cite{CLAS:2013rxx}, covering photon energies from threshold up to $11\ \text{GeV}$.
Theoretically, a number of studies based on effective hadronic Lagrangians have investigated the production mechanisms of these two reactions, spanning laboratory photon energies from threshold to about $5\ \text{GeV}$ and yielding many valuable insights~\cite{Nam:2004xt, Nam:2005jz, Nam:2005uq, Nam:2003uf, Sibirtsev:2005ns, Titov, Xie:2010yk, Nam:2008jy, Wei:2021qnc}.

Consulting the Particle Data Group (PDG) review~\cite{ParticleDataGroup:2024cfk} shows that $\pi p$ reactions serve as an ideal platform for exploring the light-hadron spectrum~\cite{WANG:2025fmh}.
Nevertheless, studies of the specific processes $\pi^{-} p \to K\Lambda(1405)$ and $\pi^{-} p \to K\Lambda(1520)$ remain scarce—a gap that motivates the present work.
We note that several earlier experiments have reported cross-section data for $\pi^{-} p \to K\Lambda(1405)$ and $\pi^{-} p \to K\Lambda(1520)$, providing a valuable basis for integrating and reanalyzing these measurements.

In this study, we investigate the production mechanisms of the hyperon excited states $\Lambda(1405)$ and $\Lambda(1520)$ in $\pi p$ scattering.
We also employ the constituent counting rule to examine the internal quark configurations of $\Lambda(1405)$ and $\Lambda(1520)$.
Furthermore, we analyze the Dalitz process $\pi^- p \to K\Lambda^* \to K\pi\Sigma$ and assess its experimental feasibility, aiming to offer useful theoretical guidance for future experimental studies.

This paper is organized as follows.
After this introduction, \cref{zhangjie2} presents the effective Lagrangians and scattering amplitudes used in our calculation.
Numerical results for the cross sections and the constituent-counting-rule analysis are given in \cref{zhangjie3}.
The Dalitz process and its experimental feasibility are discussed in \cref{zhangjie4}, followed by a brief summary in \cref{zhangjie5}.

\section{THE PRODUCTION OF $\Lambda(1405)$ AND $\Lambda(1520)$ VIA $\pi^{-} p$ reactions} \label{zhangjie2}
\begin{figure}[htbp]
	\centering
	\includegraphics[width=1.0\linewidth]{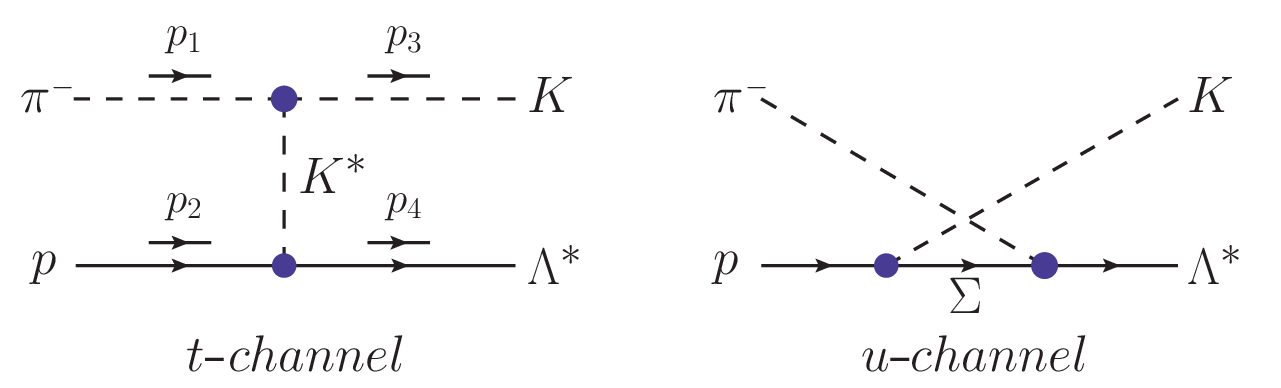}
	\caption{Feynman diagrams for $\pi^{-} p\rightarrow K\Lambda^*$.}
	\label{fmt}
\end{figure}

	The basic tree-level Feynman diagrams of $\pi^{-} p\rightarrow K\Lambda^*$ reaction, where $\Lambda^*$ stands for $\Lambda(1405)$ or $\Lambda(1520)$, are shown in \cref{fmt}.  These include $t$-channel $K^*$ exchange, and $u$-channel $\Sigma$ exchange.

To evaluate the pion-induced production of $K\Lambda(1405)$ and $K\Lambda(1520)$, the relevant Lagrangian densities are required, which have been used in~\cite{Wang:2024xvq,Wei:2021qnc,Xiang:2020phx,Wang:2024qnk,Cheng:2016ddp,Wang:2019dsi,He:2017aps,Kim:2015ita}. 
For the $t$-channel via $K^*$ exchange, the effective Lagrangians read as
\begin{eqnarray}
	\mathcal{L}_{\pi KK^*}&=&g_{\pi KK^*}[\bar K(\partial^{\mu}\bm{\tau}\cdot\bm{\pi})-(\partial^{\mu}\bar K)\bm{\tau}\cdot\bm{\pi}]K^*_{\mu} + \mathrm{H.c.},\\
	\mathcal{L}_{K^*N{\Lambda(1405)} }
	&=&g_{K^*N{\Lambda(1405)} }\bar{\Lambda}(1405)\gamma_5\gamma^{\mu}K^*_{\mu}N + \mathrm{H.c.},\\
		\mathcal{L}_{K^*N{\Lambda(1520)} }
	&=&-\frac{ig_{K^*N\Lambda(1520)}}{m_{K^*}}\bar{\Lambda}(1520)^\mu\gamma^{\nu}(\partial_\mu K^*_\nu-\partial_\nu K^*_\mu)N\notag \\
	&& + \mathrm{H.c.}.
\end{eqnarray}%
The Lagrangians corresponding to the $u$-channel are written as
\begin{eqnarray}
	\mathcal{L}_{KN{\Sigma} }
&=&g_{KN{\Sigma}}\bar{N}\gamma_5K\bm{\tau}\cdot\bm{\Sigma} + \mathrm{H.c.},\\
	\mathcal{L}_{\pi\Sigma{\Lambda(1405)}}&=&{g_{\pi\Sigma{\Lambda(1405)}}}\bar\Sigma\bm{\tau}\cdot\bm{\pi}{\Lambda(1405)} + \mathrm{H.c.},\\
	\mathcal{L}_{\pi\Sigma{\Lambda(1520)}}&=&\frac{{g_{\pi\Sigma{\Lambda(1520)}}}}{m_\pi}\bar\Sigma\gamma_{5}\bm{\tau}\cdot\partial^\mu\bm{\pi}{\Lambda(1520)_\mu} + \mathrm{H.c.},
\end{eqnarray}%
where $\bm{\tau}$ is the Pauli matrix. The symbols $\bm{\pi}$, $K$ and $K^*$ are the $\pi$, $K$ and $K^*$ meson fields, respectively. Similarly, $N$, $\Sigma$, ${\Lambda(1405)}$ and ${\Lambda(1520)}$ are the nucleon, $\Sigma$ baryon, ${\Lambda(1405)}$ and ${\Lambda(1520)}$ fields, respectively. 

The coupling constants of the $\pi KK^*$, $KN{\Sigma}$ interactions were given in many theoretical works~\cite{Wang:2024xvq,Xiang:2020phx}, and we take $g_{\pi KK^*}$ = 3.10, $g_{KN\Sigma  }$ = 2.69. We treat the coupling constants $g_{K^*N{\Lambda(1405)}}$ and $g_{K^*N{\Lambda(1520)}}$ as free parameters to be determined through fitting.

The coupling constants $g_{\pi\Sigma\Lambda(1405)}$ and $g_{\pi\Sigma\Lambda(1520)}$ can be determined from the decay widths $\Gamma_{\Lambda(1405) \rightarrow \pi\Sigma}$ and $\Gamma_{\Lambda(1520) \rightarrow \pi\Sigma}$, as detailed in ~\cite{Wang:2019dsi}
\begin{eqnarray}
	\Gamma_{\Lambda(1405) \rightarrow \pi\Sigma}=\frac{3g^2_{\pi\Sigma{\Lambda(1405)}}(m_\Sigma+E_\Sigma)}{4\pi m_{\Lambda(1405)}}|\vec{p}_\Sigma^{ c.m.}|,\\
	\Gamma_{\Lambda(1520) \rightarrow \pi\Sigma}=\frac{g^2_{\pi\Sigma{\Lambda(1520)}}(E_\Sigma-m_\Sigma)}{4\pi m^2_\pi m_{\Lambda(1520)}}|\vec{p}_\Sigma^{ c.m.}|^3,
\end{eqnarray}%
with
\begin{eqnarray}
	|\vec{p}_\Sigma^{ c.m.}|=\frac{\lambda(m^2_{\Lambda^*}, m^2_\pi, m^2_\Sigma)}{2m_{\Lambda^*}},\quad
	E_\Sigma=\sqrt{|\vec{p}_\Sigma^{ c.m.}|^2+m^2_\Sigma},
\end{eqnarray}%
where $\lambda$ is the K$\ddot{a}$llen function with a definition of $\lambda(x,y,z) =\sqrt{(x-y-z)^2-4yz}$. Moreover, $m_{\Lambda^*}$ denotes the mass of  either $\Lambda(1405)$ or $\Lambda(1520)$. The value of $g_{\pi\Sigma\Lambda(1405)}=1.58$ is determined by the decay width of $\Lambda(1405)\rightarrow \pi\Sigma$, $ \Gamma_{\Lambda(1405)\rightarrow \pi\Sigma} = 50.50 $ MeV, as advocated by PDG\cite{ParticleDataGroup:2024cfk}. In a similar manner, the coupling constant $g_{\pi\Sigma\Lambda(1520)}=2.13$ is determined from the decay width $\Gamma_{\Lambda(1520)\rightarrow \pi\Sigma} = 6.61 $ MeV given by PDG. 

For the $t$-channel, the form factor is used in our calculation as follows,
\begin{eqnarray}
 F_{t}(q_{K^*}) =\left(\frac{\Lambda{^2_{t}}-m_{K^*}^2}{\Lambda{^2_{t}}-q_{K^*}^2}\right)^2,
 \end{eqnarray}%
 where $q_{K^*}$ and $m_{K^*}$ designate the four-momentum and mass of the exchanged meson, respectively. 
 
To consider the size of the hadron, for the $u$-channel with intermediate baryon, the form factor is adopted in our calculation as follows:
\begin{eqnarray}
	F_u(q_{\Sigma}) = \frac{\Lambda{^4_u}}{\Lambda{^4_u}+(q_{\Sigma}^2-m_{\Sigma}^2)^2},
\end{eqnarray}%
where $q_{\Sigma}$ and $m_{\Sigma}$  are the four-momentum and mass of the exchanged baryon, respectively.

Based on the Lagrangians above, the scattering amplitude for the reaction $\pi^{-} p\rightarrow K\Lambda(1405)$ and $\pi^{-} p\rightarrow K\Lambda(1520)$ can be constructed as
\begin{eqnarray}
	\mathcal{M}_{\pi^{-} p\rightarrow K\Lambda(1405)}&=&\bar{u}(p_{4})(\mathcal{A}_{t}+\mathcal{A}_{u}){u}(p_{2}),\\
		\mathcal{M}_{\pi^{-} p\rightarrow K\Lambda(1520)}&=&\bar{u}_\mu(p_{4})(\mathcal{B}_{t}+\mathcal{B}_{u}){u}(p_{2}),
\end{eqnarray}
where $u_\mu$ and $u$ are dimensionless Rarita-Schwinger and Dirac spinors, respectively. 

The reduced amplitudes $\mathcal{A}_{t}$, $\mathcal{A}_{u}$ for $t$- and $u$-channel contributions read as
\begin{eqnarray}
\label{Amp1}
\mathcal{A}^{(K^*)}_{t} &=&-\sqrt{2}g_{K^*N{\Lambda(1405)}}g_{\pi KK^*}F_t(q)\gamma_{5}\gamma_{\mu}\frac{\mathcal{P}^{\mu\nu}}{{t-m_{K^*}^2}}\notag \\
&&\times(p_{1v}+p_{3v}), \\ 
\label{Amp2}
\mathcal{A}^{(\Sigma)}_{u} &=&-i2g_{\pi\Sigma\Lambda(1405)}g_{K\Sigma N}F_u(q)\frac{q\mkern-8mu/_\Sigma+m_\Sigma}{{u-m_{\Sigma}^2}}\gamma_{5}, 
\end{eqnarray}%
and the reduced amplitudes $\mathcal{B}_{t}$, $\mathcal{B}_{u}$ for $t$- and $u$-channel contributions are written as
\begin{eqnarray}
	\label{Amp33}
	\mathcal{B}^{(K^*)}_{t} &=&i\sqrt{2}\frac{g_{K^*N{\Lambda(1520)}}g_{\pi KK^*}}{m_{K^*}}F_t(q)\gamma^{\sigma}\Big[(p_1-p_3)_\mu g_{\sigma\nu}\notag \\
	&&-(p_1-p_3)_\sigma g_{\mu\nu}\Big]\frac{\mathcal{P}^{\mu\nu}}{{t-m_{K^*}^2}} (p_{1}^{\mu}+p_{3}^{\mu}), \\ 
	\label{Amp4}
	\mathcal{B}^{(\Sigma)}_{u} &=&2\frac{g_{\pi\Sigma\Lambda(1520)}g_{K\Sigma N}}{m_\pi}F_u(q)\gamma_{5}p_1^\mu
	\frac{q\mkern-8mu/_\Sigma+m_\Sigma}{{u-m_{\Sigma}^2}}\gamma_{5},  
\end{eqnarray}%
with
\begin{eqnarray}
    \mathcal{P}^{\mu\nu}&=&i(-g^{\mu\nu}+q_{K^*}^\mu q_{K^*}^\nu/{m_{K^*}^2}),
\end{eqnarray}%
where $s=(p_1+p_2)^2$, $u =(p_1- p_4)^2$, and $t =(p_1- p_3)^2$ are the Mandelstam
variables.

The Regge trajectory model has proven to be successful
in analyzing hadron production at intermediate and high energies ~\cite{Wang:2024qnk,Wang:2024xvq,Wang:2015hfm,Ozaki:2009wp, Wang:2015zcp,Wang:2017qcw,Storrow:1983ct,Xiang:2020phx}. In this model, the Reggeization can be done by replacing the $t$-channel propagator in the Feynman amplitudes  (Eq.(\ref{Amp1}, \ref{Amp33})) with the Regge propagator
\begin{eqnarray}
\frac{1}{t-m_{K^*}^2}\rightarrow\left(\frac{s}{s_{scale}}\right)^{\alpha_{K^*}(t)-1}\frac{\pi\alpha^{\prime }_{K^*}}{\Gamma[{\alpha_{K^*}(t)}]\sin[\pi{\alpha_{K^*}(t)}]}.
\end{eqnarray}%
 The scale factor $s_{scale}$ is fixed at 1 GeV. In addition, the Regge trajectory of $\alpha_{K^*}(t)$ is written as ~\cite{Ozaki:2009wp},
\begin{eqnarray}
\alpha_{K^*}(t)=1 + 0.85(t - m_{K^*}^2).
\end{eqnarray}%
It can be seen that no free parameters have been added after
the introduction of the Regge model.
\section{Numerical results}\label{zhangjie3}
After the above preparation, one can calculate the differential cross sections of the reactions $\pi^{-} p\rightarrow K\Lambda(1405)$ and $\pi^{-} p\rightarrow K\Lambda(1520)$ and combine the experimental data for correlation analysis. The differential cross section in the center of mass (c.m.) frame is given by:
\begin{eqnarray}
\label{Aamp}
\frac{d\sigma}{dcos\theta} = \frac{1}{32\pi s}\frac{|\vec{p}_3^{c.m.}|}{|\vec{p}_1^{c.m.}|}\left(\frac{1}{2}\sum\limits_\lambda|\mathcal{M}|^2 \right),
\end{eqnarray}%
where $\theta$ represents the angle between the
outgoing $K$ meson and the direction of the $\pi$ beam
in the center of mass frame. $\vec{p}_1^{c.m.}$ and $\vec{p}_3^{c.m.}$ represent the three-momenta of the initial $\pi$ beam and the final $K$ meson, respectively.
\subsection{Results for $K\Lambda(1405)$ Production}
With the help of the MINUIT code in the CERNLIB, the experimental data~\cite{Dahl:1967pg,Crennell:1972km,Thomas:1973uh} of $\pi^{-} p\rightarrow K\Lambda(1405)$ reaction will be fitted in this work. The experimental data for the total cross section and the differential cross section lead us to use the $\chi^2$ fitting algorithm to determine the values of the free parameters. The fitting parameters, along with a $\chi^2/\mathrm{d.o.f.}=2.343$, are presented in Table \ref{1}. 

\begin{table}[h]
	\renewcommand\arraystretch{1.8} 
	\caption{Fitted values of free parameters based on all experimental data from Refs. ~\cite{Dahl:1967pg,Crennell:1972km,Thomas:1973uh}.}
	\label{1}{\footnotesize \centering
	\setlength{\tabcolsep}{3.65mm}{
		\begin{tabular}{cccc}
			\toprule[1pt]\toprule[1pt]
			$\Lambda_t$&$\Lambda_u$& $g_{K^*N{\Lambda(1405)}}$&	$\chi^2/\mathrm{d.o.f.}$  \\
			\hline
			1.856$\pm$0.051&1.345$\pm$0.056&2.476$\pm$0.370&2.343\\
			\bottomrule[1pt]\bottomrule[1pt]
		\end{tabular}}
	}
\end{table}
\begin{figure}[htbp]
	\centering
	\includegraphics[width=1.0\linewidth]{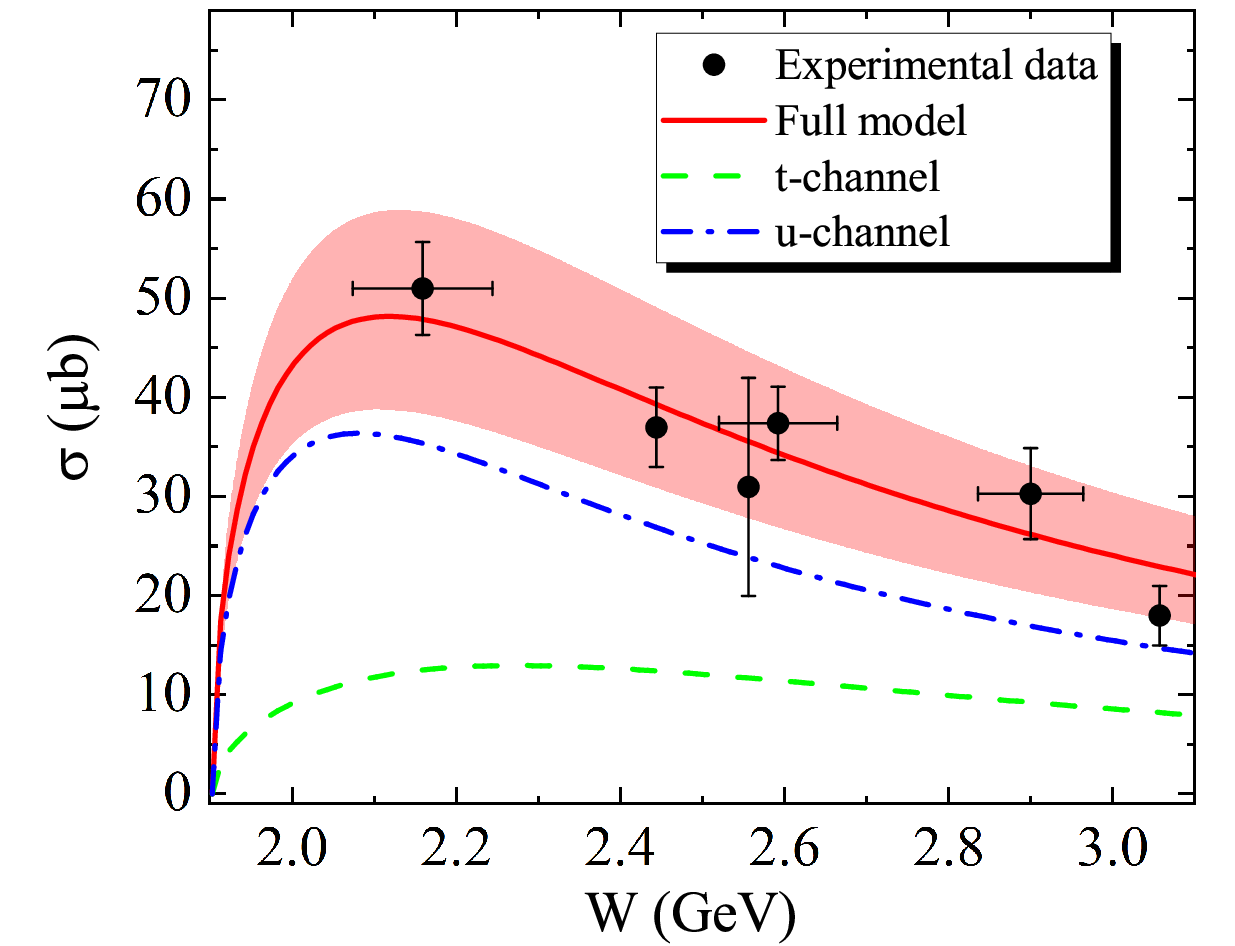}
	\caption{The total cross section for the reaction $\pi^{-} p\rightarrow K\Lambda(1405)$. The band stands for the error bar of the three fitting parameters in Table \ref{1}.}
	\label{tcs1}
\end{figure}
\begin{figure}[b]
	\centering
	\includegraphics[width=1.0\linewidth]{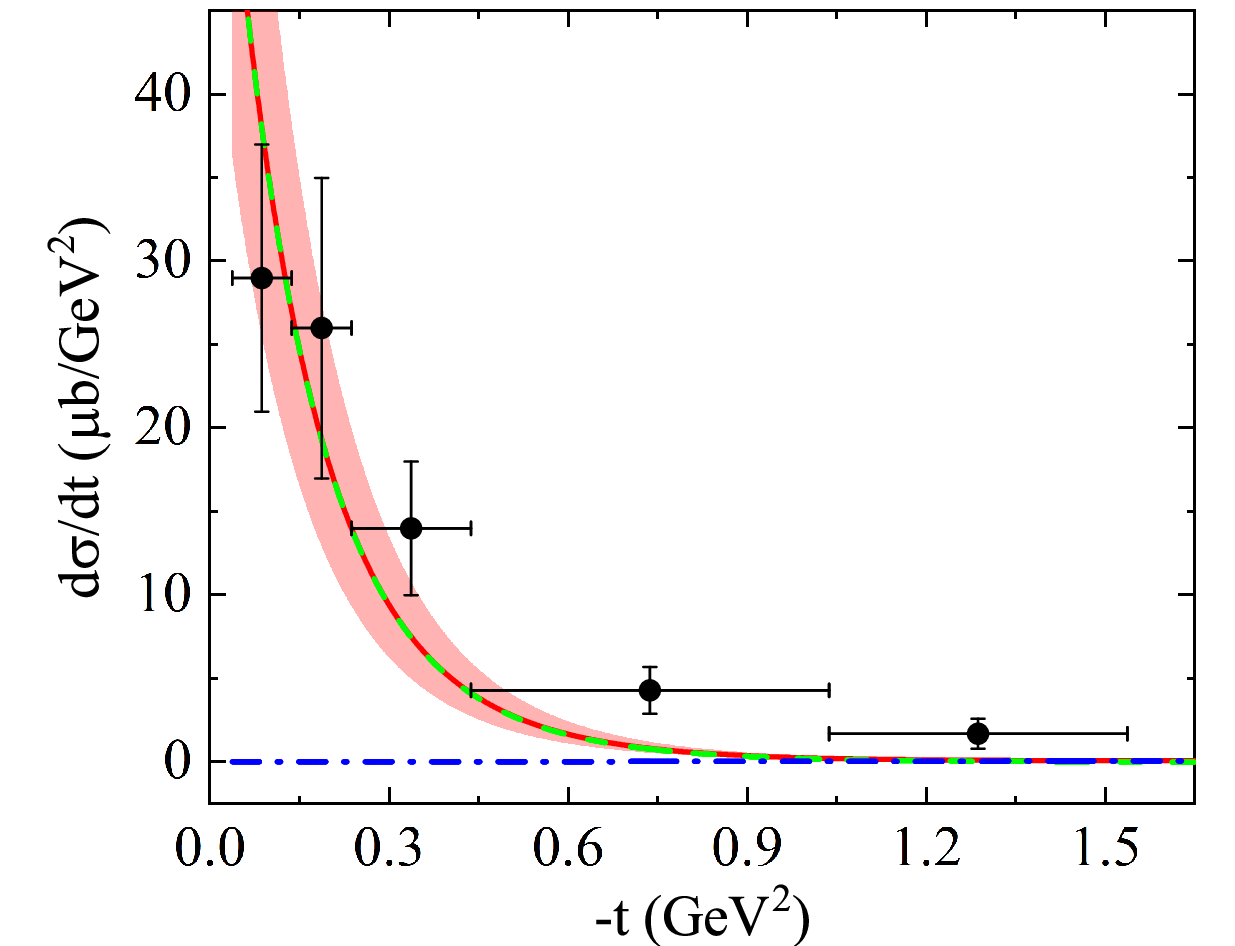}
	\caption{The $t$-distribution for the reaction $\pi^{-} p\rightarrow K\Lambda(1405)$ at a center of mass energy of $W = 3.057$ GeV. The experimental data are from Ref. ~\cite{Crennell:1972km}. Here, the notation is the same as in \cref{tcs1}.
	}
	\label{t1}
\end{figure}
\begin{figure}[htbp]
	\centering
	\includegraphics[width=1.0\linewidth]{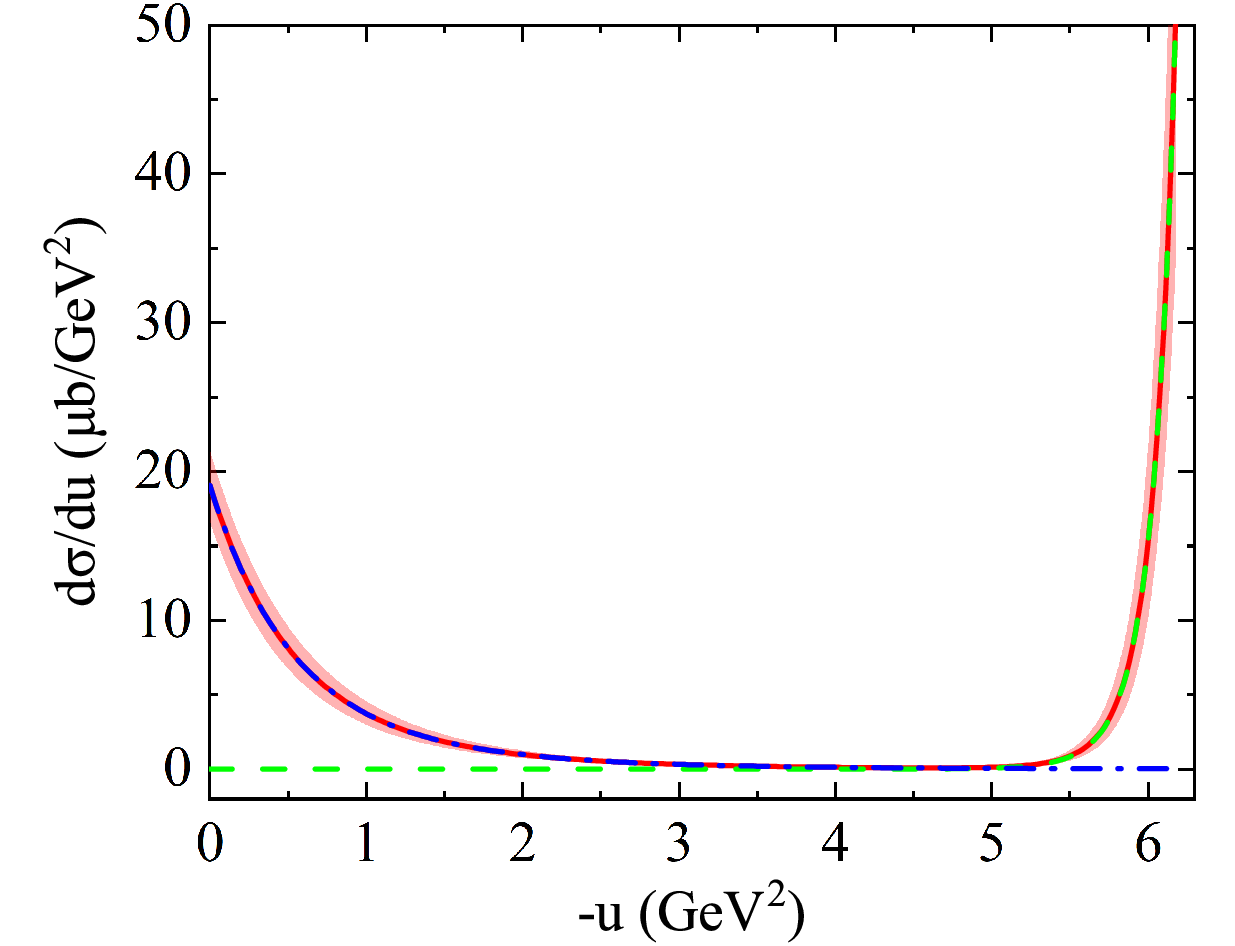}
	\caption{The $u$-distribution for the reaction $\pi^{-} p\rightarrow K\Lambda(1405)$ at a center of mass energy of $W = 3.057$ GeV. Here, the notation is the same as in \cref{tcs1}.
	}
	\label{u1}
\end{figure}
In \cref{tcs1}, we present the total cross section of the  $\pi^{-} p\rightarrow K\Lambda(1405)$ reaction. As depicted in \cref{tcs1}, the total cross section exhibits a clear variation trend within the energy range of $W=1.90$ to $3.10$ GeV. Notably, there is a  peak between $W=2.00$ to $2.20$ GeV, indicating the potential for observing the $\Lambda(1405)$ resonance through $\pi^-p$ interactions within this energy range. \cref{t1} and \cref{u1} show the $t$-distribution and $u$-distribution  of the $\pi^{-} p\rightarrow K\Lambda(1405)$ reaction at $W = 3.057$ GeV. It can be observed that as $|t|$ increases and $|u|$ decreases, the contribution from the $t$-channel becomes less significant, while that from the $u$-channel grows increasingly prominent. Notably, it is evident from \cref{tcs1} and \cref{t1} that both the total cross section and the differential cross section of the $t$-distribution can give a reasonable description of the experimental
data. In \cref{cos1}, we show the differential cross sections of the $\pi^-p \to K\Lambda(1405)$ reaction computed at various center-of-mass energies. Analysis of these results reveals a clear angular dependence: the $t$-channel contribution dominates at forward angles, while the $u$-channel contribution becomes prevalent at backward angles. This angular division is further quantified in \cref{t3}, which illustrates that for the $\pi^-p \to K\Lambda(1405)$ reaction across different center-of-mass energies, the $u$-channel contribution grows increasingly significant with increasing $t$, whereas the $t$-channel contribution diminishes gradually. In \cref{cos3}, we present the differential cross section of the $\pi^-p\rightarrow K\Lambda(1405)$ reaction as a function of center-of-mass energies when $\cos\theta=1$. It can be seen from this figure that the differential cross section continuously increases as the center-of-mass energy increases, and the shapes of the cross sections of the $u$-channel and the $t$-channel differ significantly.
\begin{figure}[htbp]
	\centering
	\includegraphics[width=1.0\linewidth]{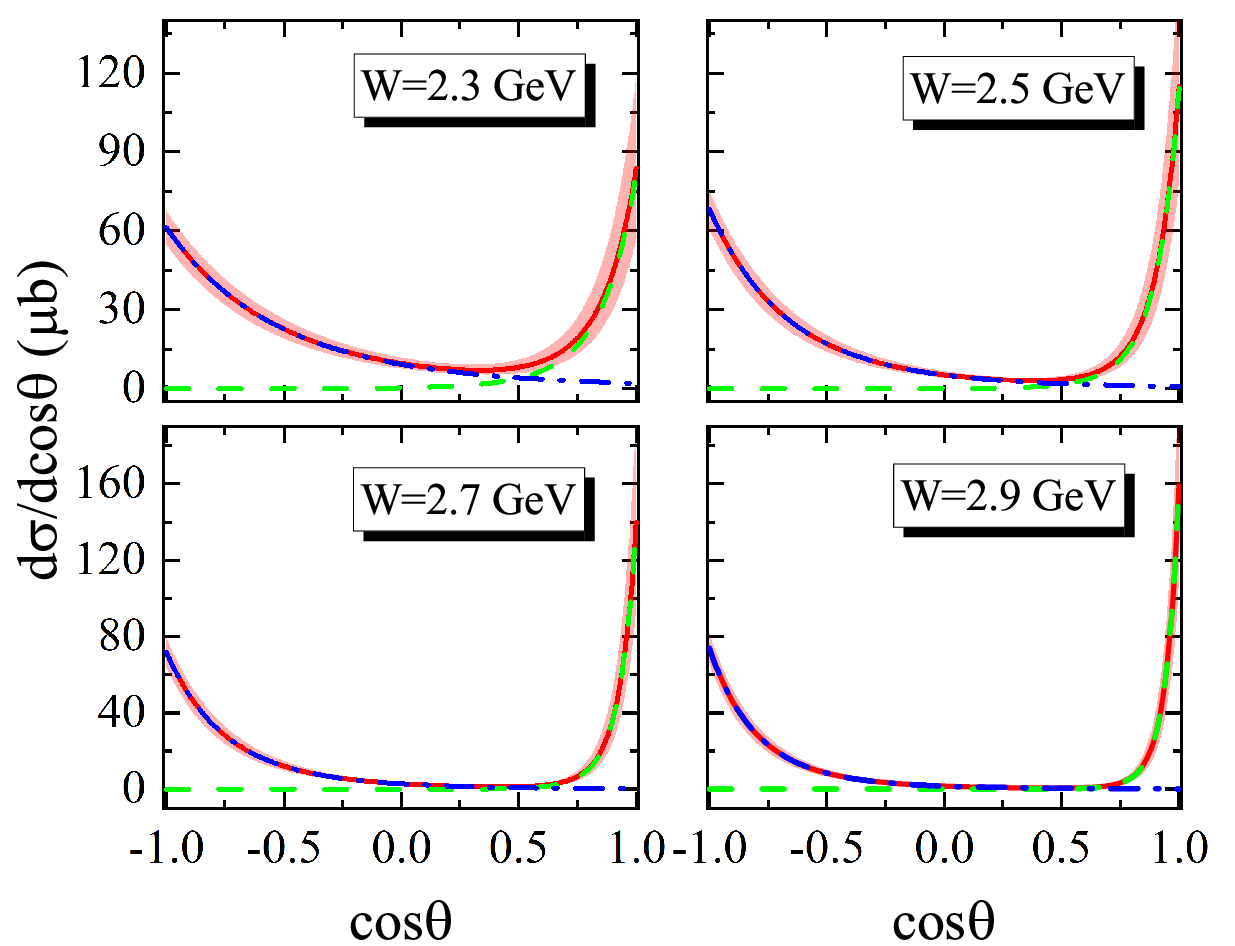}
	\caption{The differential cross section $d \sigma/d \cos\theta$ of the $\pi^{-} p\rightarrow K\Lambda(1405)$ process as a function of $\cos\theta$ at different c.m. energies.  Here, the notation is the same as in \cref{tcs1}.}
	\label{cos1}
\end{figure}

\begin{figure}[b]
	\centering
	\includegraphics[width=1.0\linewidth]{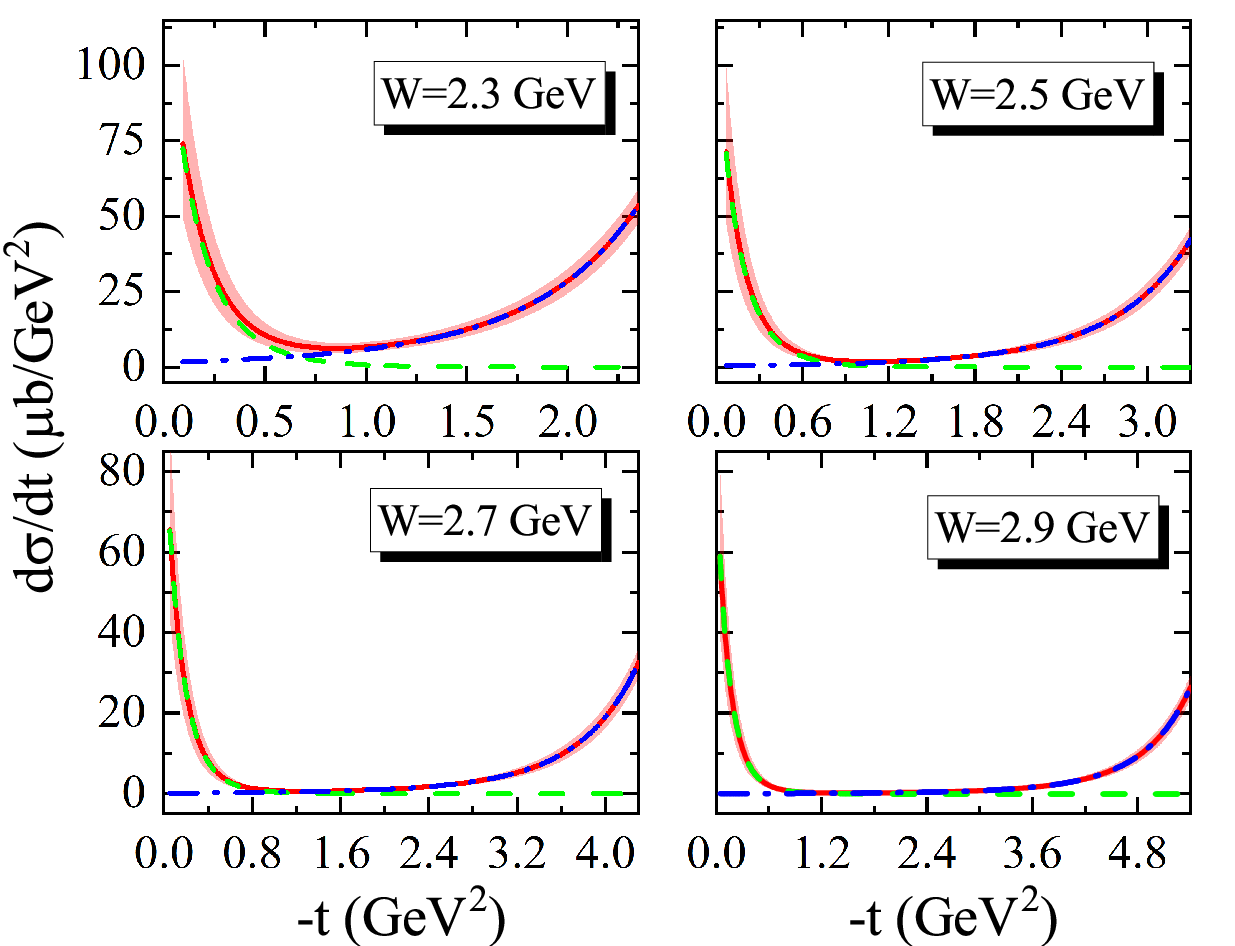}
	\caption{The $t$-distribution for the $\pi^{-} p\rightarrow K\Lambda(1405)$ reaction at different c.m. energies. Here, the notation is the same as in \cref{tcs1}. }
	\label{t3}
\end{figure}
\begin{figure}[htbp]
	\centering
	\includegraphics[width=1.0\linewidth]{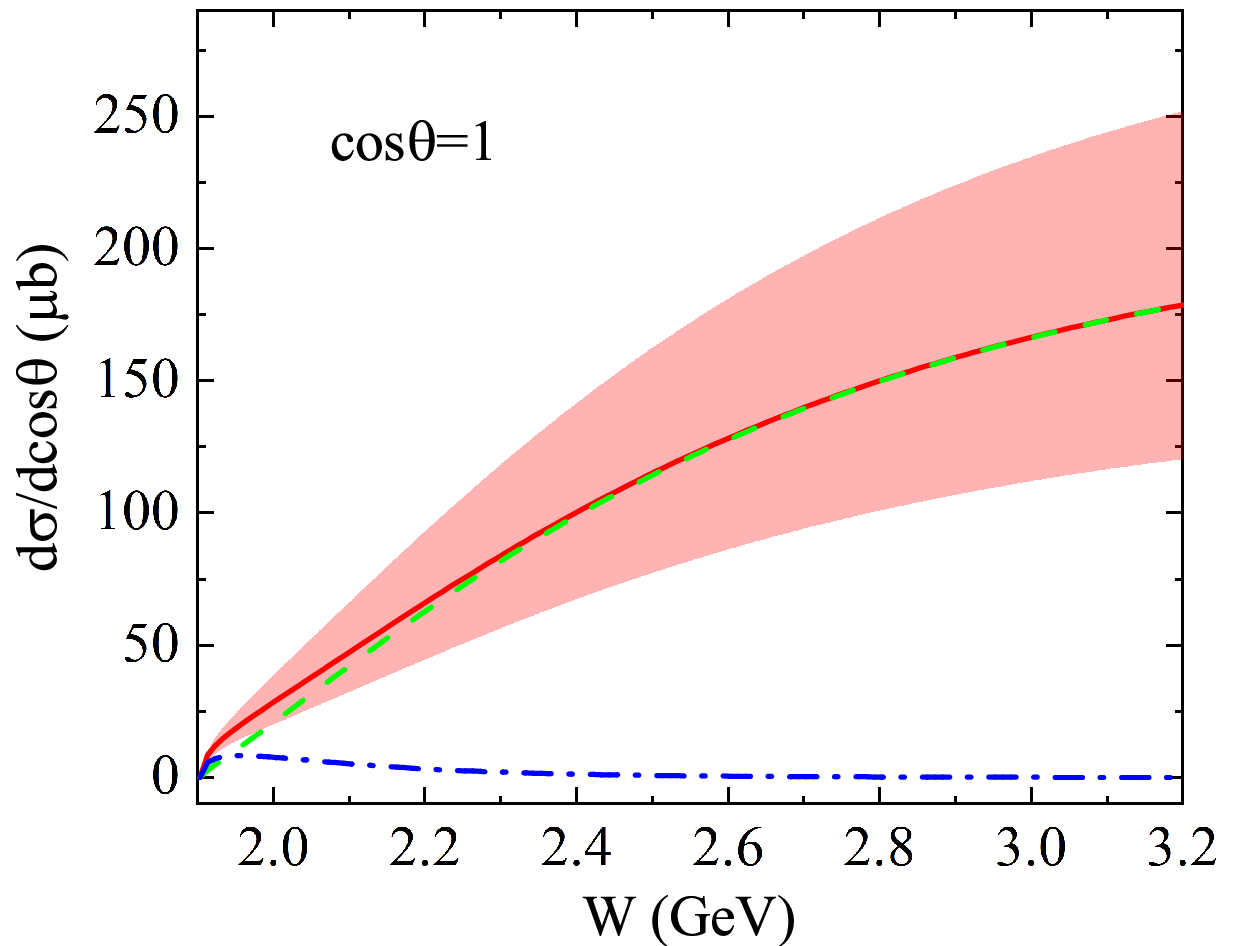}
	\caption{The differential cross section d$\sigma$/dcos$\theta$ for the $\pi^-p\rightarrow K\Lambda(1405)$ reaction varies with different c.m. energies when $\cos\theta=1$. Here, the notation is the same as in \cref{tcs1}.}
	\label{cos3}
\end{figure}

\subsection{Results for $K\Lambda(1520)$ Production}
\begin{figure}[htbp]
	\centering
	\includegraphics[width=1.0\linewidth]{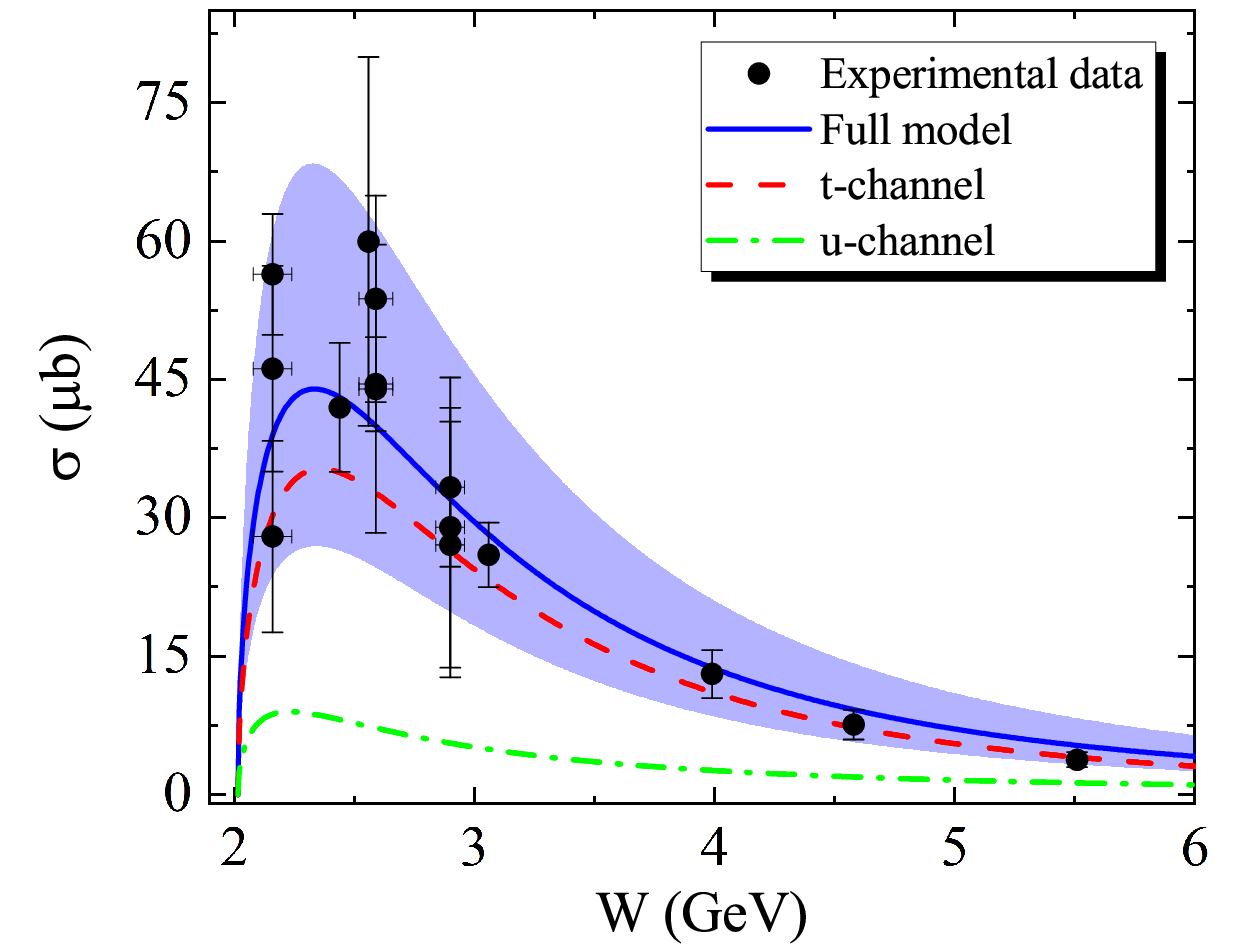}
	\caption{The total cross section for the reaction $\pi^-p\rightarrow K\Lambda(1520)$. The band stands for the error bar of the three fitting parameters in Table \ref{free}.}
	\label{tcs2}
\end{figure}

\begin{figure}[htbp]
	\centering
	\includegraphics[width=1.0\linewidth]{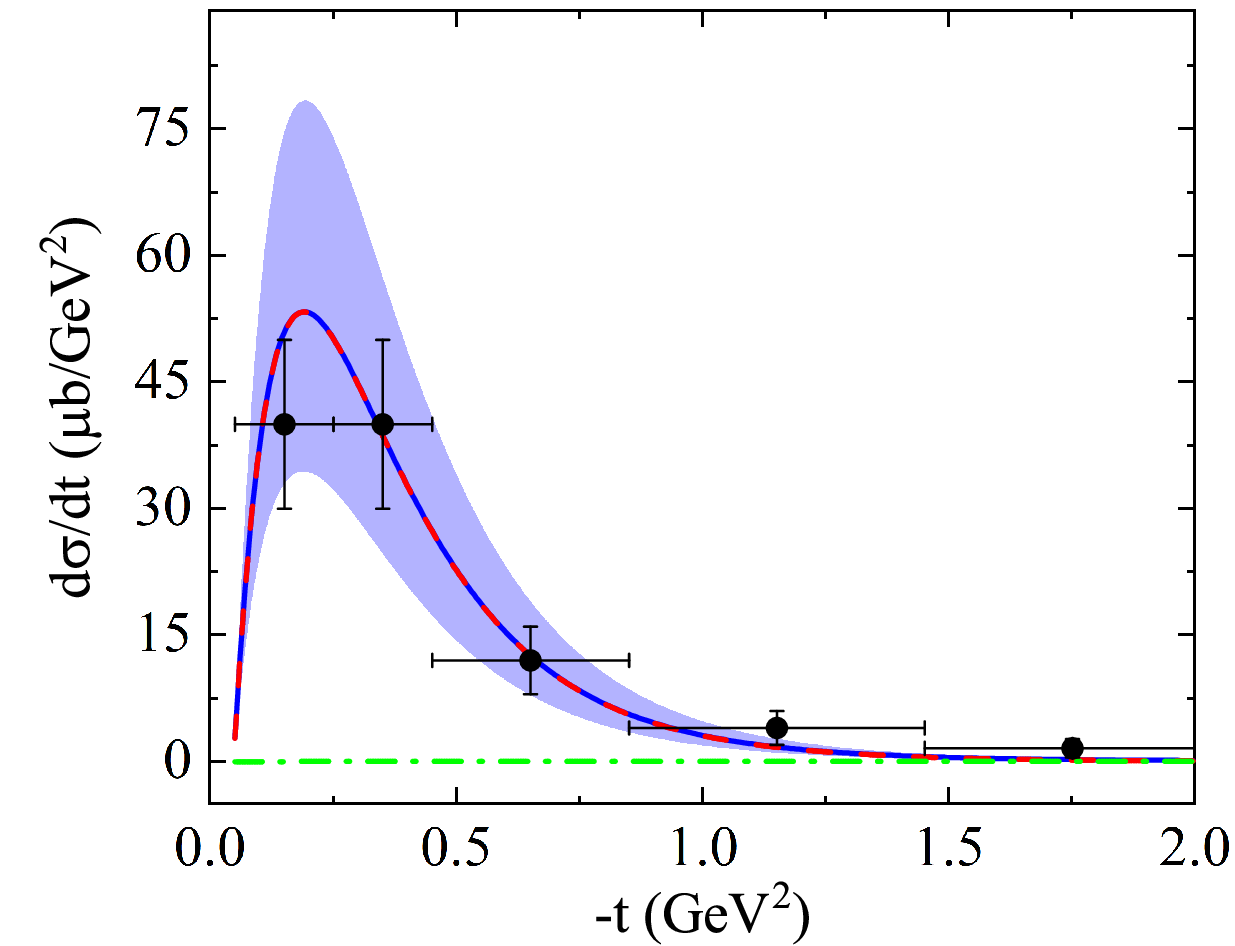}
	\caption{The $t$-distribution for the reaction $\pi^-p\rightarrow K\Lambda(1520)$ at a center of mass energy of $W = 3.057$ GeV. The experimental data are from Ref. ~\cite{Crennell:1972km}.  Here, the notation is the same as that in \cref{tcs2}.}
	\label{t2}
\end{figure}

\begin{figure}[b]
	\centering
\includegraphics[width=1.0\linewidth]{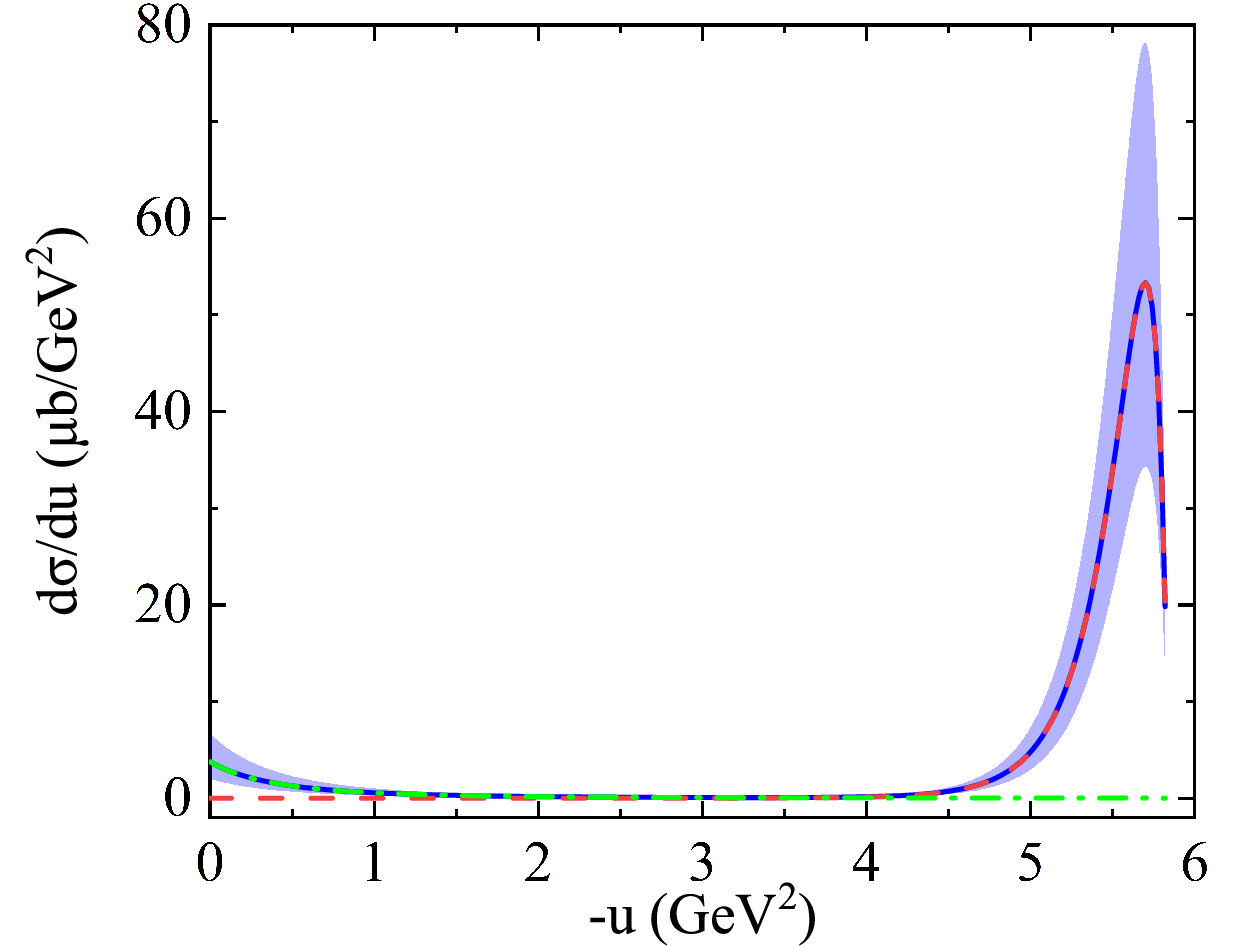}
	\caption{The $u$-distribution for the reaction $\pi^-p\rightarrow K\Lambda(1520)$ at a center of mass energy of $W = 3.057$ GeV.  Here, the notation is the same as that in \cref{tcs2}.}
	\label{u2}
\end{figure}

\begin{figure}[htbp]
	\centering
	\includegraphics[width=1.0\linewidth]{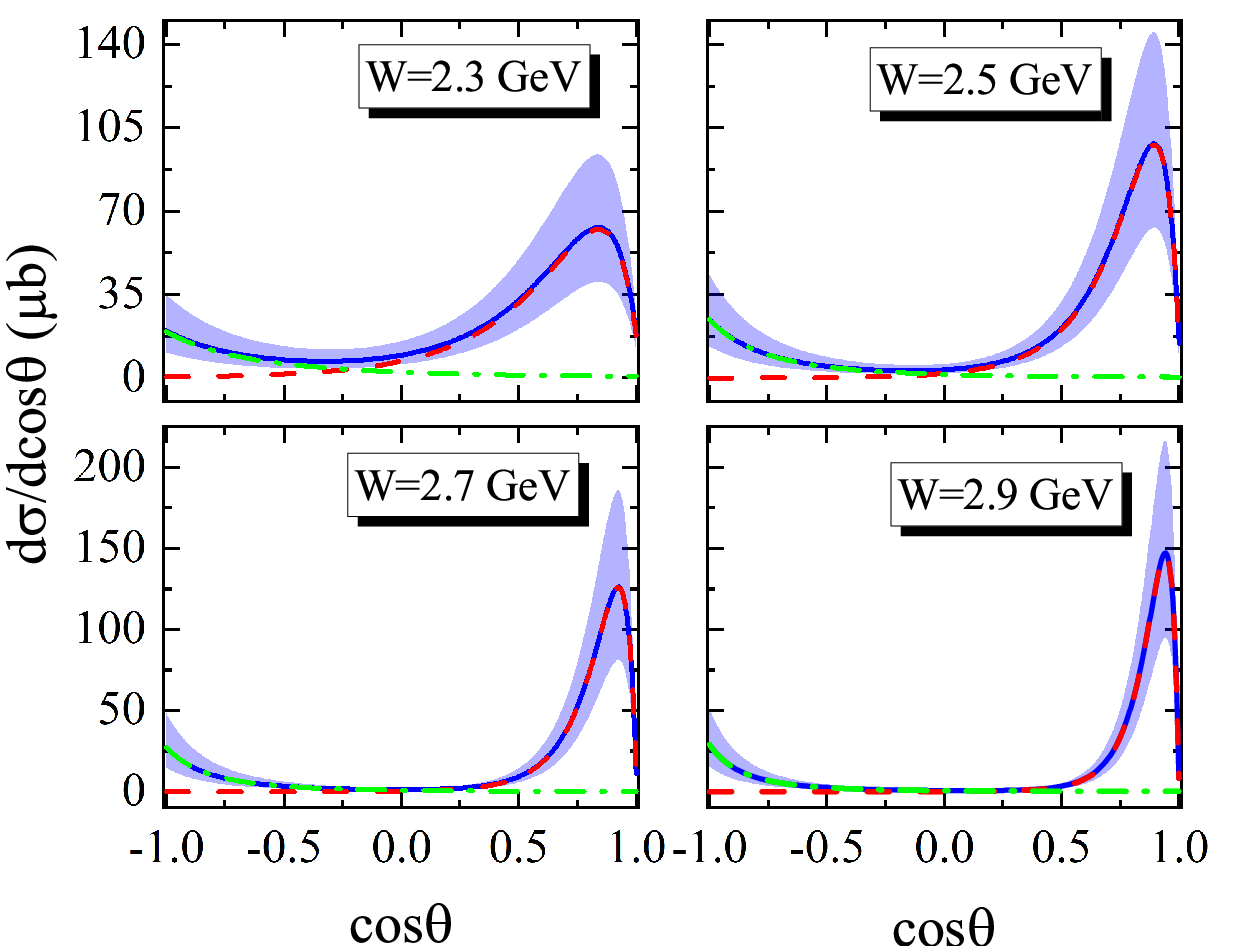}
	\caption{The differential cross section $d \sigma/d \cos\theta$ of the $\pi^-p\rightarrow K\Lambda(1520)$ process as a function of $\cos\theta$ at different c.m. energies. Here, the notation is the same as that in \cref{tcs2}.}
	\label{cos2}
\end{figure}

\begin{figure}[b]
	\centering
	\includegraphics[width=1.0\linewidth]{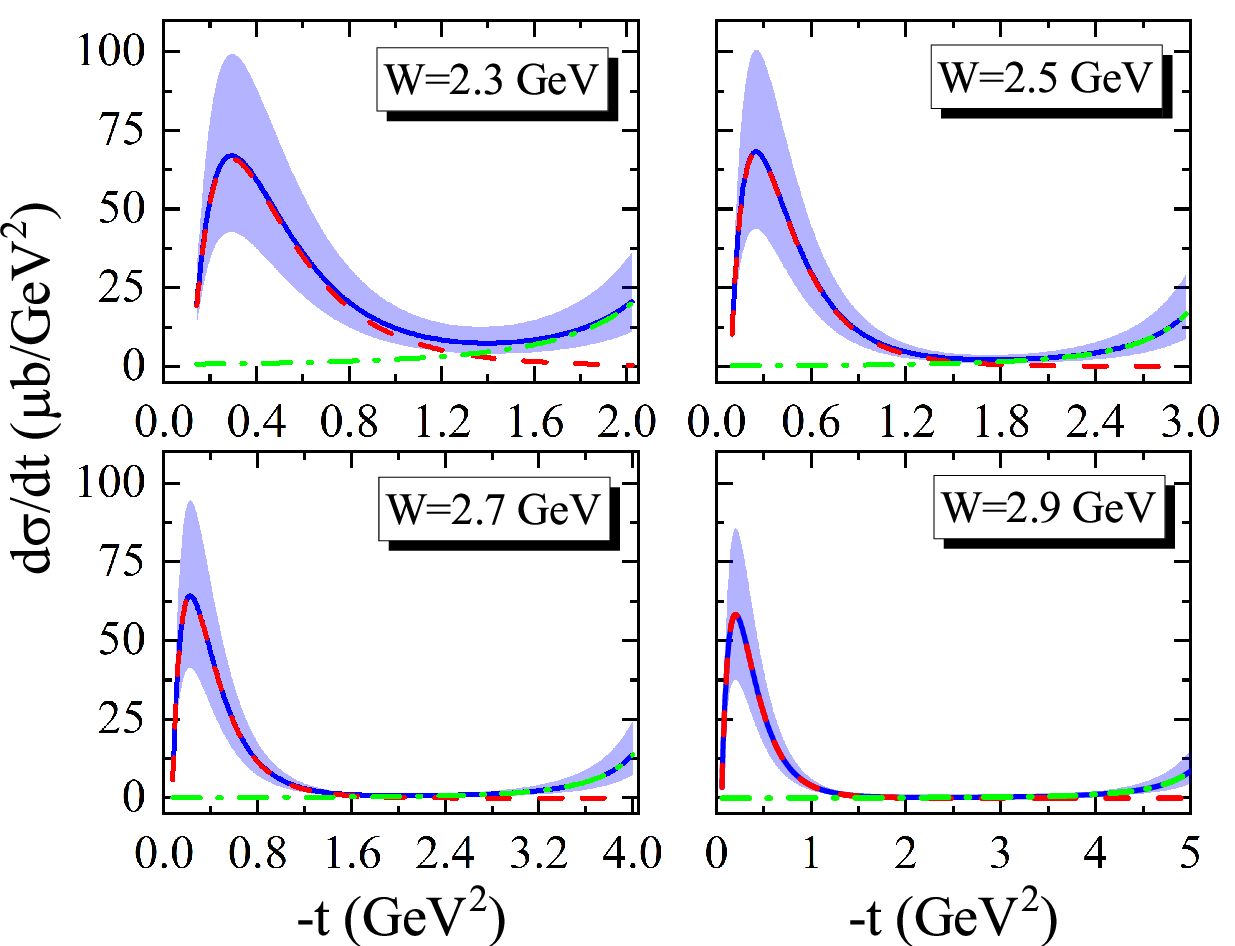}
	\caption{The $t$-distribution for the $\pi^-p\rightarrow K\Lambda(1520)$ reaction at different c.m. energies. Here, the notation is the same as in \cref{tcs2}. }
	\label{t4}
\end{figure}

\begin{figure}[htbp]
	\centering
	\includegraphics[width=1.0\linewidth]{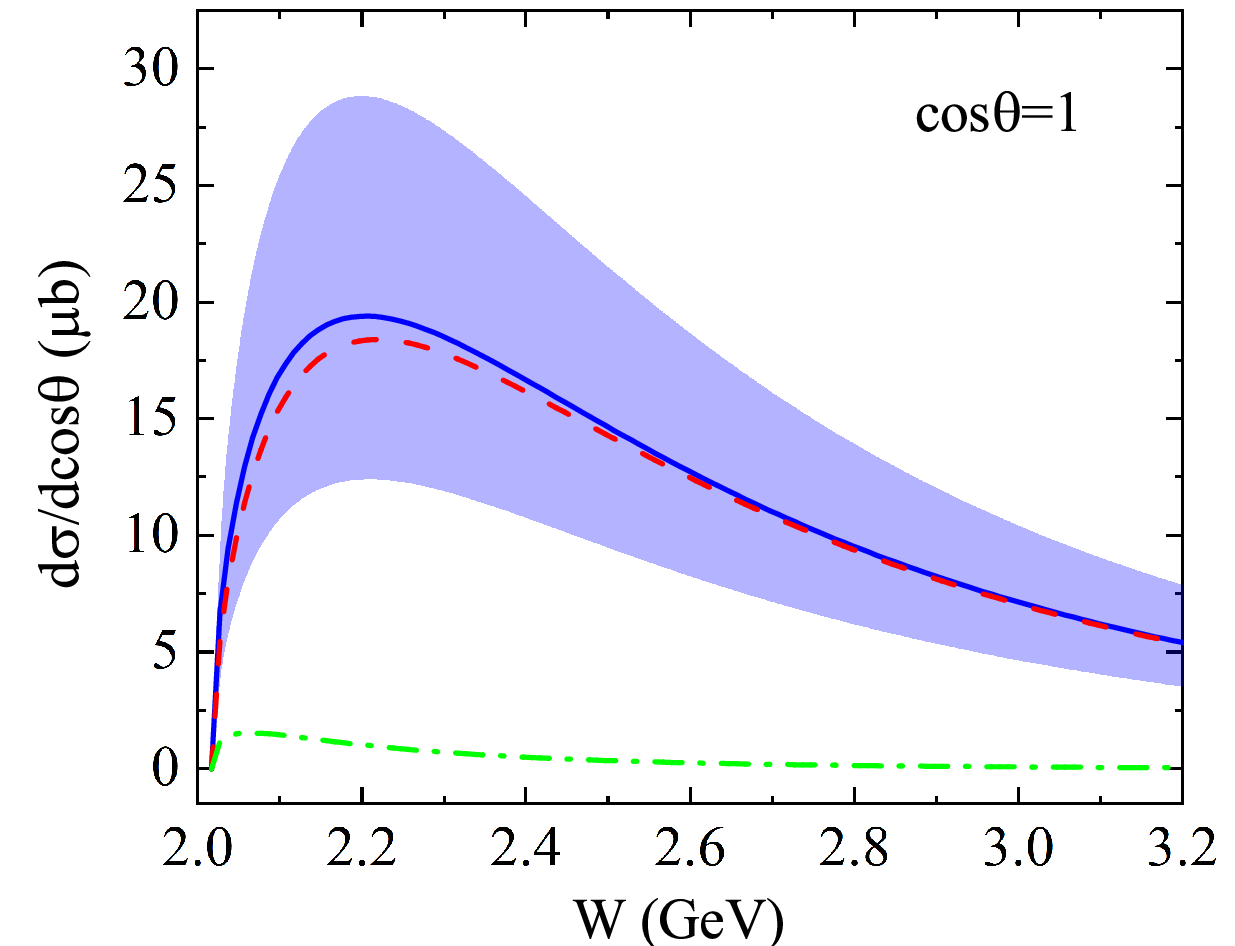}
	\caption{The differential cross section d$\sigma$/dcos$\theta$ for the $\pi^-p\rightarrow K\Lambda(1520)$ reaction varies with different c.m. energies when $\cos\theta=1$. Here, the notation is the same as that in \cref{tcs2}.}
	\label{cos4}
\end{figure}

Similar to the case of calculating $\pi^-p\rightarrow K\Lambda(1405)$, we also calculate the cross section of $\pi^-p\rightarrow K\Lambda(1520)$ as shown in \cref{tcs2,t2,u2,cos2,cos4,t4}. We use the same fitting scheme as Table \ref{1}, treating the coupling constant $g_{K^*N\Lambda{1520}}$ and the cutoffs $\Lambda_t$ and $\Lambda_u$ as free parameters. The fitting parameters involved are shown in Table \ref{free} and the result of considering the full model gives the value of $\chi^2$/d.o.f. = 1.259. 

As depicted in \cref{tcs2}, the total cross section for the $\pi^-p\rightarrow K\Lambda(1520)$ reaction exhibits a prominent peak within the energy range of $W=2.20$ to $2.50$ GeV, which suggests the potential for observing the $\Lambda(1520)$ resonance via $\pi^-p$ interactions within this energy interval. \cref{t2} and \cref{u2} show the $t$- and $u$-distributions of the $\pi^- p \to K\Lambda(1520)$ reaction at $W = 3.057$ GeV; their shapes differ from those of the $t$- and $u$-distributions for the $\pi^- p \to K\Lambda(1405)$ reaction (\cref{t1} and \cref{u1}) at the same energy. In \cref{cos2}, the differential cross sections of the $\pi^{-} p\rightarrow K\Lambda(1520)$ reaction at different center-of-mass energies are calculated. \cref{t4} shows the $t$-distribution for the $\pi^{-} p\rightarrow K\Lambda(1520)$ reaction at different center-of-mass energies. As $t$ increases, the contribution of the $u$-channel becomes increasingly significant, while the contribution of the $t$-channel gradually decreases. In \cref{cos4}, we present the differential cross section of the $\pi^-p\rightarrow K\Lambda(1520)$ reaction as a function of center-of-mass energies when $\cos\theta=1$. Notably, a prominent peak is observed in the energy range of $W=2.10$ to $2.30$ GeV, indicating that the $\Lambda(1520)$ resonance may be accessible through $\pi^-p$ interactions within this energy interval.
\begin{table}[h]
	\renewcommand\arraystretch{1.8} 
	\caption{Fitted values of free parameters based on all experimental data from Refs. ~\cite{Dahl:1967pg,Crennell:1972km,Thomas:1973uh}.}
	\label{free}{\footnotesize \centering
		\setlength{\tabcolsep}{3.65mm}{
			\begin{tabular}{cccc}
				\toprule[1pt]\toprule[1pt]
				$\Lambda_t$&$\Lambda_u$& $g_{K^*N{\Lambda(1520)}}$&	$\chi^2/\mathrm{d.o.f.}$  \\
				\hline
		   	    1.488$\pm$0.036&0.514$\pm$0.040&13.200$\pm$1.883&1.259\\
				\bottomrule[1pt]
    \bottomrule[1pt]
		\end{tabular}}
	}
\end{table}



\subsection{Internal Structure of Hadrons by Constituent Counting Rule}
The constituent counting rule was proposed by Stanley J. Brodsky in 1973 and developed in the decade subsequent to the creation of  perturbative QCD~\cite{Lepage:1980fj,Mueller:1981sg,Lepage:1979zb,Brodsky:1974vy,Brodsky:1973kr,Matveev:1973ra,Dong:2019stt}. In essence, it stands for the conformality and scale invariance of QCD at high energies and thus is applicable to a wide range of field theories. For example, it has been derived nonperturbatively in AdS/QCD~\cite{Polchinski:2001tt}. The constituent counting rule indicates that in the exclusive process $a + b \rightarrow c + d$ with large momentum transfer, quarks should share large momenta so that they can stick together to form a hadron by exchanging hard gluons. And the cross section of the $a + b\rightarrow c + d$ exclusive reaction should scale like $d\sigma/dt \sim s^{2-n}f(\theta_{\mathrm{c.m.}})$ with $n = n_a +n_b +n_c +n_d$, where $n_i$ is the number of constituents in the particle $i$, $s$ and $t$ are Mandelatam variables, and $\theta_{\mathrm{c.m.}}$ is the scattering angle in the center-of-mass system. Notably, since the factor $n_h$ clearly indicates the internal configuration in the hadron, this scaling relation can be used for finding internal configurations of exotic hadrons. The constituent counting rule was confirmed by two-body hadronic exclusive reactions at BNL (Brookhaven National Laboratory)-AGS (Alternating Gradient Synchrotron)~\cite{Baller:1988tj}. In general, the constituent counting rule can well describe the differential cross section distribution behavior of many hadron-hadron scattering processes in the large momentum transfer region~\cite{JeffersonLabHallA:2002vjd,JeffersonLabHallA:2004gxp,White:1994tj,Amaryan:2021cnj}. In our previous work, we calculated the cross sections of $\pi^-p\rightarrow K^*\Sigma$ at $\theta_{\mathrm{c.m.}}=90^\circ$, by fitting the experimental data and the numerical results to the expression $d\sigma/dt = (\text{constant}) \times s^{2-n}$, We found that the fitted values of $n$ are very close to 10, confirming the
accuracy of the theoretical prediction for the cross section.~\cite{Wang:2024xvq}.

$\Lambda(1520)$ has long been regarded as a three-quark baryon; therefore, the number of internal quark configurations it contains should be $3$, leading to the scaling behavior of the differential cross section as $d\sigma/dt \sim 1/s^8$. The cross section at $\theta_{\mathrm{c.m.}}=90^{\circ}$ was calculated for $\Lambda(1520)$ production processes $\pi^-p \rightarrow K\Lambda(1520)$, with the results presented in \cref{33}. Furthermore, by fitting the numerical results shown in \cref{33} with the expression $d\sigma/dt=({\rm constant})\times s^{2-n}$, we present the obtained scaling factors $n$ in \cref{zfjsgz}.  We find that for the reaction of $\pi^-p\rightarrow K\Lambda(1520)$, the fitted values are consistent with $n=10$, namely with the assignment $ n_{\Lambda(1520)}=3$, confirming that the theoretical cross section prediction is reliable.

Furthermore, the constituent counting rule can be employed to explore the internal structure of exotic hadron candidates. Therefore, we calculated the cross section of the $\Lambda(1405)$ production process $\pi^-p \rightarrow K\Lambda(1405)$ at a center-of-mass scattering angle of $\theta_{\mathrm{c.m.}}=90^\circ$, aiming to investigate the internal structure of $\Lambda(1405)$. If $\Lambda(1405)$ is a five-quark hadron, the total number of constituents is $n = 2 + 3 + 2 + 5 = 12$, and the differential cross section obeys the scaling law $d\sigma/dt \sim 1/s^{10}$; whereas if $\Lambda(1405)$ is an ordinary three-quark baryon, the total number of constituents is $n = 2+ 3+ 2 + 3= 10$, with the corresponding scaling behavior of the differential cross section given by $d\sigma/dt \sim 1/s^8$.

\begin{figure}[htbp]
	\centering
	\includegraphics[width=1.0\linewidth]{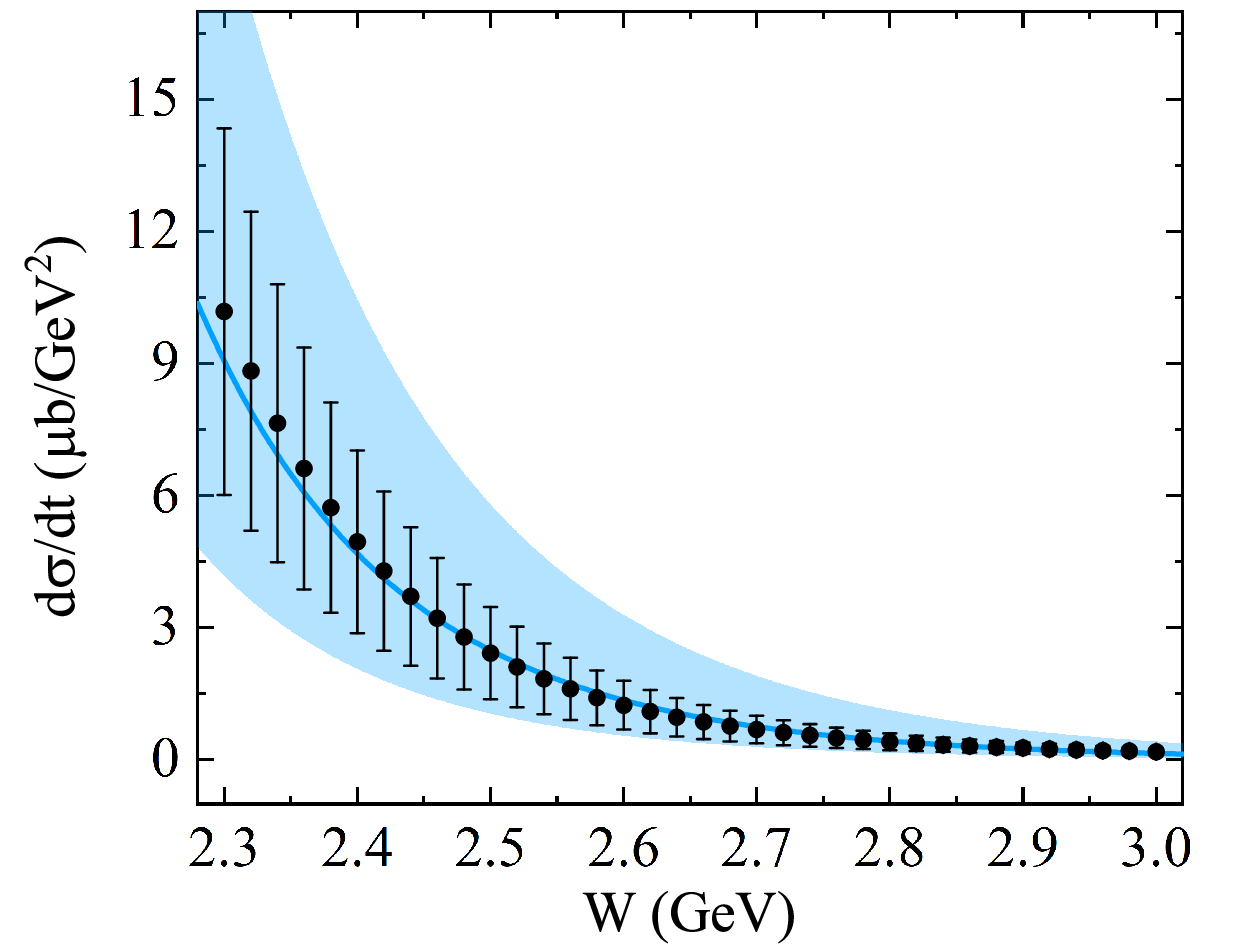}
	\caption{Differential cross section of $\pi^-p\rightarrow K\Lambda(1520)$, $d\sigma/dt$, at large meson production angle $\theta=90^\circ$ in c.m. as a function of $(W=\sqrt{s})$. The numerical points are obtained from \cref{33}. The band stands for the error bar of the fitting parameters in \cref{zfjsgz}.}
	\label{ccr1520}
\end{figure}

\begin{figure}[htbp]
	\centering
	\includegraphics[width=1.0\linewidth]{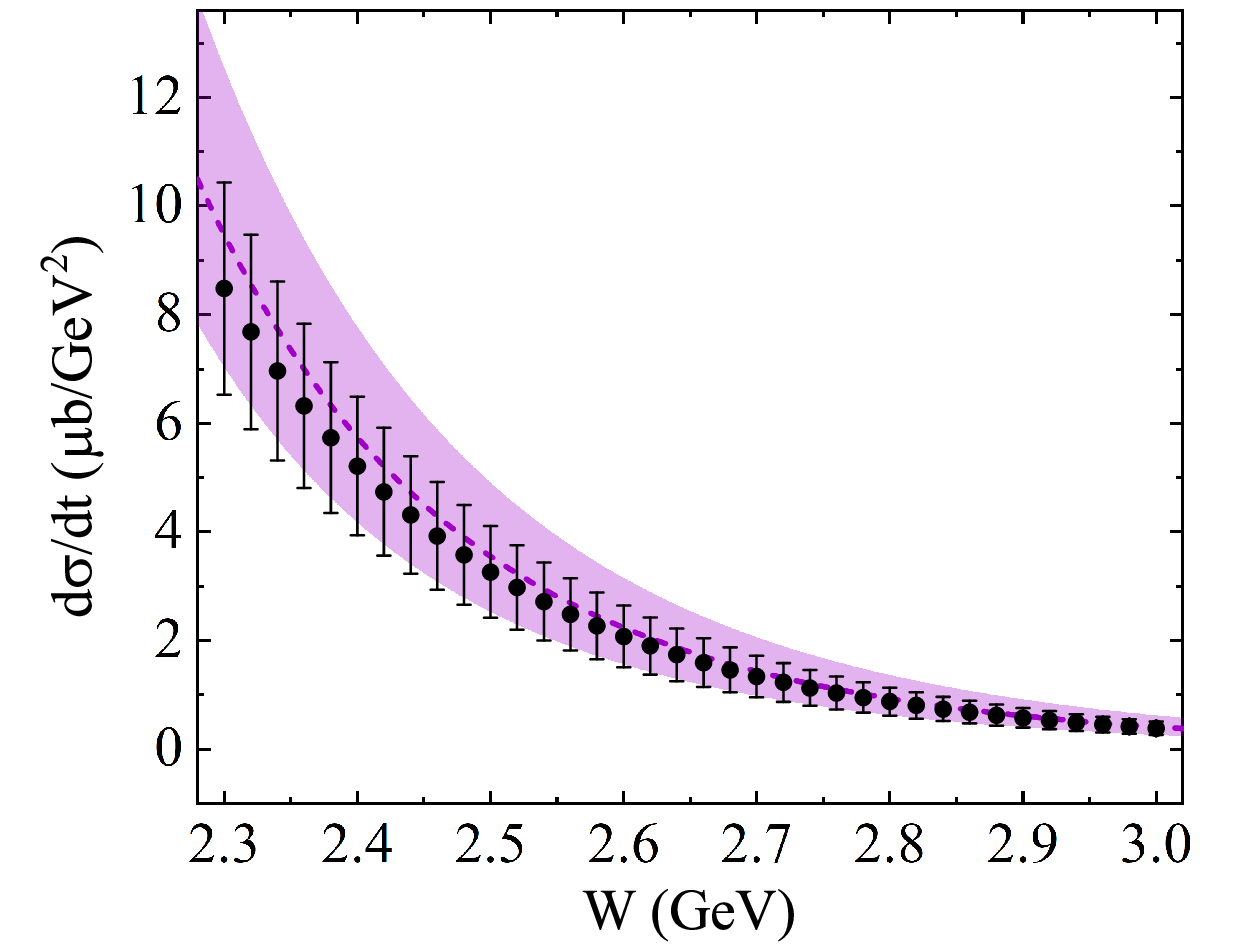}
	\caption{Differential cross section of $\pi^-p\rightarrow K\Lambda(1405)$, $d\sigma/dt$, at large meson production angle $\theta=90^\circ$ in c.m. as a function of $(W=\sqrt{s})$. The numerical points are obtained from \cref{32}. The band stands for the error bar of the fitting parameters in \cref{zfjsgz}.}
	\label{ccr1405}
\end{figure}

However, it can be seen from the results in \cref{zfjsgz} that for the reaction $\pi^-p\rightarrow K\Lambda(1405)$, the fitted values of $n$ are close to 8, which are significantly lower than the theoretically expected values of $n=12$ or $n=10$. We speculate that the significant influence of unquenched effects on the hyperon excited state $\Lambda(1405)$ may be responsible for this particular result. \cref{ccr1405} and \cref{ccr1520} show the differential cross sections, $d\sigma/dt$, at large meson production angle ($\theta=90^\circ$ in c.m.) as a function of $\sqrt{s}$ for $\pi^-p\rightarrow K\Lambda(1405)$ and $\pi^-p\rightarrow K\Lambda(1520)$, respectively.

Notably, there is currently no experimental data on the $t$-distribution cross section at $\cos\theta=0$. The data points shown in \cref{32} are the results of our theoretical calculations, which is very unfavorable for us to obtain an accurate value of $n$. From our analysis, we propose that experimentalists measure the differential cross section $d\sigma/dt$ of the $\pi^-p\rightarrow K\Lambda(1405)$ reaction at $\cos\theta = 0$ for different center-of-mass energies to draw a solid conclusion for the scaling factor $n$. It is especially important for clarifying the internal constituents of an exotic hadron candidate $\Lambda(1405)$.
\begin{table}[h]
	\renewcommand\arraystretch{1.5} 
	\caption{The cross sections of $\pi^-p\rightarrow K\Lambda(1520)$ at $\theta_{\mathrm{c.m.}}=90^\circ$ in the energy range of 2.3–3.0 GeV.}
	\label{33}{\footnotesize \centering
		\setlength{\tabcolsep}{3.5mm}{
			\begin{tabular}{cccc}
				\toprule[1pt]\toprule[1pt]
				$\sqrt{s}~(\rm GeV)$ & $\frac{d\sigma}{dt}~(\mu b/{\rm GeV}^2)$& $\sqrt{s}~({\rm GeV})$&$\frac{d\sigma}{dt}~(\mu b/{\rm GeV}^2)$ \\
				
				\hline
    2.30&10.175$\pm$4.157&2.66&0.862$\pm$0.389\\
				
				2.32&8.825$\pm$3.622&2.68&0.768$\pm$0.348\\
				
				2.34&7.645$\pm$3.153&2.70&0.687$\pm$0.313\\
				
				2.36&6.617$\pm$2.743&2.72&0.617$\pm$0.282\\
				2.38&5.724$\pm$2.385&2.74&0.556$\pm$0.255\\
				2.40&4.952$\pm$2.074&2.76&0.502$\pm$0.232\\
				2.42&4.285$\pm$1.805&2.78&0.455$\pm$0.211\\
				2.44&3.710$\pm$1.571&2.80&0.414$\pm$0.192\\
				2.46&3.216$\pm$1.369&2.82&0.378$\pm$0.176\\
				2.48&2.789$\pm$1.195&2.84&0.346$\pm$0.162\\
				2.50&2.424$\pm$1.044&2.86&0.318$\pm$0.149\\
				2.52&2.111$\pm$0.915&2.88&0.292$\pm$0.137\\
				2.54&1.842$\pm$0.803&2.90&0.270$\pm$0.127\\
				2.56&1.611$\pm$0.706&2.92&0.249$\pm$0.117\\
				2.58&1.413$\pm$0.623&2.94&0.231$\pm$0.109\\
				2.60&1.243$\pm$0.551&2.96&0.214$\pm$0.101\\
				2.62&1.097$\pm$0.489&2.98&0.199$\pm$0.094\\
				2.64&0.971$\pm$0.435&3.00&0.186$\pm$0.088\\
				\bottomrule[1pt]\bottomrule[1pt]
		\end{tabular}}
	}
\end{table}
\begin{table}[h]
	\renewcommand\arraystretch{1.5} 
	\caption{The cross sections of $\pi^-p\rightarrow K\Lambda(1405)$ at $\theta_{\mathrm{c.m.}}=90^\circ$ in the energy range of 2.3–3.0 GeV.}
	\label{32}{\footnotesize \centering
		\setlength{\tabcolsep}{3.5mm}{
			\begin{tabular}{cccc}
				\toprule[1pt]\toprule[1pt]
				$\sqrt{s}~(\rm GeV)$ & $\frac{d\sigma}{dt}~(\mu b/{\rm GeV}^2)$& $\sqrt{s}~({\rm GeV})$&$\frac{d\sigma}{dt}~(\mu b/{\rm GeV}^2)$ \\
				\hline
				2.30&8.490$\pm$1.954&2.66&1.602$\pm$0.448\\
				2.32&7.691$\pm$1.792&2.68&1.469$\pm$0.414\\
				2.34&6.973$\pm$1.645&2.70&1.348$\pm$0.383\\
				2.36&6.327$\pm$1.510&2.72&1.237$\pm$0.355\\
				2.38&5.745$\pm$1.388&2.74&1.137$\pm$0.328\\
				2.40&5.221$\pm$1.276&2.76&1.045$\pm$0.304\\
				2.42&4.748$\pm$1.174&2.78&0.960$\pm$0.282\\
				2.44&4.321$\pm$1.081&2.80&0.884$\pm$0.261\\
				2.46&3.936$\pm$0.996&2.82&0.813$\pm$0.242\\
				2.48&3.587$\pm$0.918&2.84&0.749$\pm$0.224\\
				2.50&3.272$\pm$0.846&2.86&0.690$\pm$0.208\\
				2.52&2.986$\pm$0.780&2.88&0.636$\pm$0.193\\
				2.54&2.727$\pm$0.720&2.90&0.586$\pm$0.179\\
				2.56&2.492$\pm$0.664&2.92&0.541$\pm$0.166\\
				2.58&2.278$\pm$0.614&2.94&0.499$\pm$0.154\\
				2.60&2.085$\pm$0.567&2.96&0.461$\pm$0.143\\
				2.62&1.908$\pm$0.524&2.98&0.426$\pm$0.133\\
				2.64&1.748$\pm$0.484&3.00&0.394$\pm$0.123\\
				\bottomrule[1pt]\bottomrule[1pt]
		\end{tabular}}
	}
\end{table}

\begin{table}[h]
	\renewcommand\arraystretch{1.8} 
	\caption{Fitted values of the scaling factor n using the $\chi^2$ fitting algorithm.}
	\label{zfjsgz}{\footnotesize \centering
		\setlength{\tabcolsep}{5mm}{
			\begin{tabular}{ccc}
				\toprule[1pt]\toprule[1pt]		 
				 &  $\pi^-p\rightarrow K\Lambda(1520)$&  \\
				 \hline
				 $\sqrt{s}$ &$n$& $\chi^2/\mathrm{d.o.f.}$ \\
				  \hline
				   $2.3-2.8$ GeV  &10.291$\pm$0.363&1.147\\
				$2.3-3.0$ GeV  &9.747$\pm$0.531&1.368\\
				\hline
		 
				&  $\pi^-p\rightarrow K\Lambda(1405)$&  \\
				 \hline
     $\sqrt{s}$ &$n$& $\chi^2/\mathrm{d.o.f.}$ \\
				  \hline
				$2.3-2.8$ GeV &7.568$\pm$0.347&1.180\\
				  $2.3-3.0$ GeV  &7.871$\pm$0.249&1.202\\
				\bottomrule[1pt]\bottomrule[1pt]	
		\end{tabular}}
	}
\end{table}
\section{Dalitz process and experimental feasibility}\label{zhangjie4}
\begin{figure}[b]
	\centering
\includegraphics[width=1.0\linewidth]{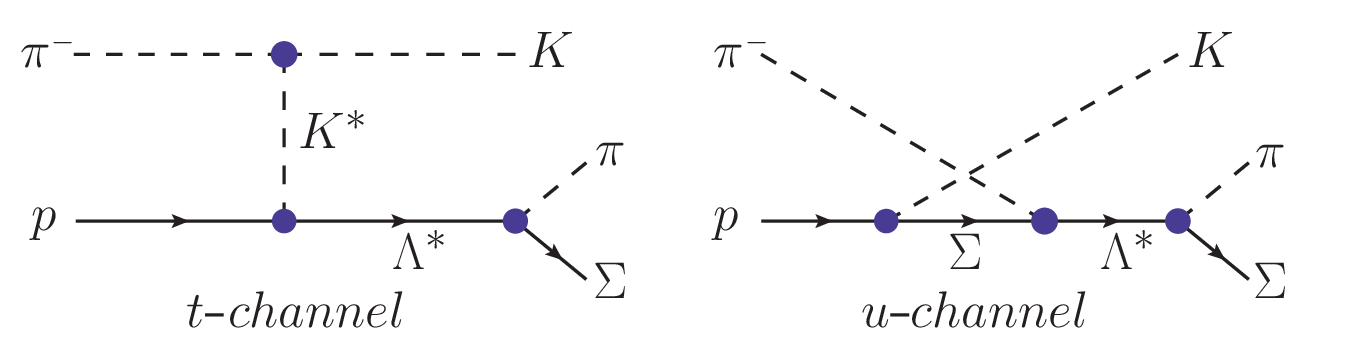}
	\caption{Feynman diagrams for the $\pi^-p\rightarrow  K\pi\Sigma$ reaction.}
	\label{dalitz}
\end{figure}
\begin{figure}[t]
	\centering
	\includegraphics[width=1.0\linewidth]{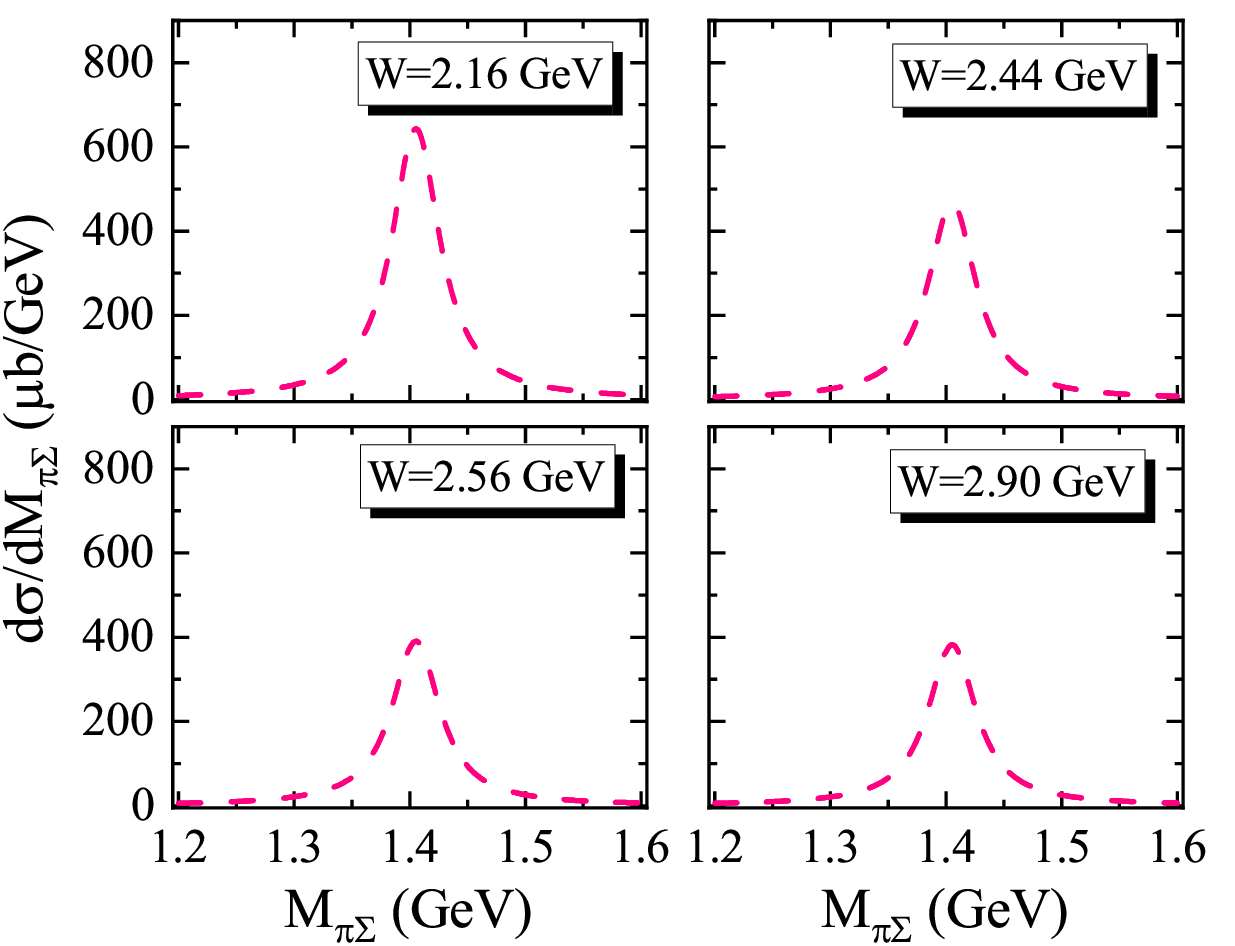}
	\caption{The invariant mass distribution $d\sigma_{\pi^-p\rightarrow 
			K\Lambda(1405) \rightarrow K\pi\Sigma}$ /$dM_{\pi\Sigma}$ at different c.m. energies $W$ = 2.16 GeV, 2.44 GeV, 2.56 GeV, and 2.90 GeV.}
	\label{dalitz1}
\end{figure}
\begin{figure}[htbp]
	\centering
	\includegraphics[width=1.0\linewidth]{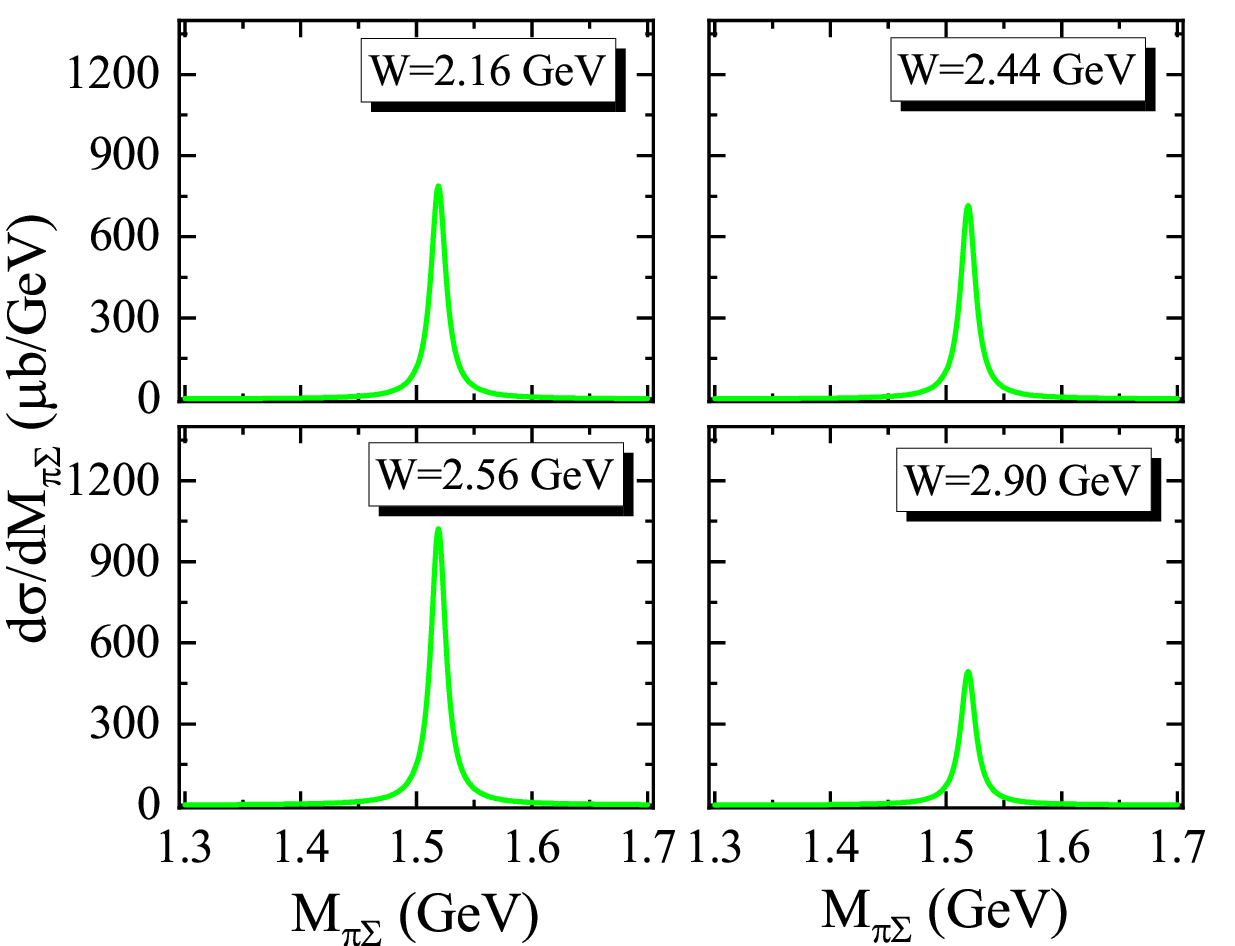}
	\caption{The invariant mass distribution $d\sigma_{\pi^-p\rightarrow 
			K\Lambda(1520) \rightarrow K\pi\Sigma}$ /$dM_{\pi\Sigma}$ at different c.m. energies $W$ = 2.16 GeV, 2.44 GeV, 2.56 GeV, and 2.90 GeV.}
	\label{dalitz2}
\end{figure}

In $\pi^-p\rightarrow K\Lambda(1405)$ and $\pi^-p\rightarrow K\Lambda(1520)$ processes, the hyperon excited states $\Lambda(1405)$ and $\Lambda(1520)$, which cannot be directly measured experimentally, must be determined by reconstructing the final particles resulting from their decay. Considering that the branching ratio of $\Lambda(1405)$ decay to $\pi\Sigma$
can reach 100.00\%~\cite{ParticleDataGroup:2024cfk} and the branching ratio of $\Lambda(1520)$ decay to $\pi\Sigma$
can reach 42.00\%~\cite{ParticleDataGroup:2024cfk}, it is necessary to analyze the Dalitz process for $\pi^-p \rightarrow K\Lambda^*\rightarrow K\pi\Sigma$, which will provide useful information for experimental measurements. In general, the invariant mass spectrum of the Dalitz process is defined based on the two-body process~\cite{Kim:2017nxg,Wang:2024qnk,Wang:2024xvq,Wang:2023lia}
\begin{eqnarray}
	\frac{d\sigma_{\pi^-p\rightarrow 
			K\Lambda^* \rightarrow K\pi\Sigma}}{dM_{\pi\Sigma}} &\approx&\frac{2M_{\Lambda^*}M_{\pi\Sigma}}{\pi}\notag\\
	&&\times\frac{\sigma_{\pi^-p\rightarrow 
			K\Lambda^*} \Gamma_{\Lambda^*\rightarrow \pi\Sigma}}{(M^2_{\pi\Sigma}-M^2_{\Lambda^*})^2+M^2_{\Lambda^*}\Gamma^2_{\Lambda^*}},
\end{eqnarray}%
where $\Gamma_{\Lambda(1405)}$= 50.50 MeV and $\Gamma_{\Lambda(1405)\rightarrow \pi\Sigma}$ = 50.50 MeV represent the total width and decay component width of $\Lambda(1405)$, respectively.  $\Gamma_{\Lambda(1520)}$= 15.73 MeV and $\Gamma_{\Lambda(1520)\rightarrow \pi\Sigma}$ = 6.61 MeV represent the total width and decay component width of $\Lambda(1520)$, respectively. Based on these parameters, we calculate the invariant mass distributions $d\sigma_{\pi^-p\rightarrow	K\Lambda(1405) \rightarrow K\pi\Sigma}/dM_{\pi\Sigma}$ and $d\sigma_{\pi^-p\rightarrow K\Lambda(1520) \rightarrow K\pi\Sigma}/dM_{\pi\Sigma}$ for center-of-mass energies ranging from $W= 2.16$ GeV to $W= 2.90$ GeV. The calculation results are exhibited in Figs. \ref{dalitz1} and \ref{dalitz2}.

It is observed from \cref{dalitz1} that peak values are observed at $M_{\pi\Sigma}$ = 1405 MeV and the peaks are not less than 381.97 $\mu b$/GeV. It has direct implications for the experimental detection of the $\Lambda(1405)$. To further investigate the feasibility of experimentally searching for the $\Lambda(1405)$ by $\pi^-p$ scattering, we should calculate the ratio $\sigma(\pi^-p \rightarrow K\Lambda(1405)\rightarrow K\pi\Sigma)/\sigma(\pi^-p\rightarrow K\pi\Sigma)$. In Ref.~\cite{Goussu:1966ps}, the experimental data for $\pi^-p\rightarrow K\pi\Sigma$ is reported at $W=1.973$ GeV. Based on this value, the ratio at $W=1.973$ GeV can be calculated as follows:
\begin{eqnarray}
	\frac{\sigma({\pi^-p\rightarrow 
			K\Lambda(1405) \rightarrow K\pi\Sigma})}{\sigma({\pi^-p\rightarrow K\pi\Sigma})} \approx 60.52\%,
\end{eqnarray}%
which further indicates that the total cross section and event number of the $\pi^-p \rightarrow K\Lambda(1405)$ process reconstructed via $\pi^-p\rightarrow K\pi\Sigma$ satisfy the requirements for experimental measurement. Figure \ref{dalitz2} shows that the peak values appear at $M_{\pi\Sigma}$ = 1520 MeV and the peaks are not less than 494.34 $\mu b$/GeV, which means that it is feasible to reconstruct the $\pi^-p\rightarrow K\Lambda(1520)$ process by the $\pi^-p \rightarrow K\pi\Sigma$ process experimentally. 


\section{Discussions and conclusions} \label{zhangjie5}

This study investigates the production mechanisms of the hyperon excited states $\Lambda(1405)$ and $\Lambda(1520)$ through the reactions $\pi^{-} p \rightarrow K\Lambda(1405)$ and $\pi^{-} p \rightarrow K\Lambda(1520)$, using an effective Lagrangian approach combined with the Reggeized trajectory model. Specifically, we consider $K^*$ meson exchange in the $t$-channel and $\Sigma$ exchange in the $u$-channel. By fitting the total cross section and $t$-distribution data for $\pi^-p\rightarrow K\Lambda^*$, distinct resonance peaks are observed: the $\pi^-p\rightarrow K\Lambda(1405)$ process peaks in the center-of-mass energy range $W=2.00$–$2.20$ GeV, while $\pi^-p\rightarrow K\Lambda(1520)$ peaks between $W=2.20$ and $2.50$ GeV. These features suggest that $\Lambda(1405)$ and $\Lambda(1520)$ can be effectively studied via $\pi^-p$ interactions within these energy intervals. The fitting results indicate that the $t$-channel dominates at forward angles and low momentum transfer ($t$), whereas the $u$-channel contributes significantly at backward angles and larger $t$.

The constituent counting rule provides critical insight into the internal structure. For $\pi^{-} p \rightarrow K\Lambda(1520)$, the extracted scaling exponent $n$ is consistent with 10, supporting $\Lambda(1520)$ as a conventional three-quark baryon. In contrast, for $\pi^{-} p \rightarrow K\Lambda(1405)$, the fitted $n$ is close to 8, which deviates notably from the theoretical expectation of 10 for a three-quark baryon and 12 for a five-quark configuration. However, the absence of experimental $t$-distribution data near $\cos\theta = 0$ limits the accuracy in determining $n$. Furthermore, given the large branching ratio of $\Lambda^{*}\rightarrow\pi\Sigma$, we analyze the Dalitz process $\pi^{-} p \rightarrow K \Lambda^{*} \rightarrow K \pi \Sigma$. The results confirm the feasibility of detecting $\Lambda^{*}$ through this cascade decay.

High-precision measurements of meson–nucleon scattering are now possible at facilities such as AMBER~\cite{Adams:2018pwt}, J-PARC~\cite{Aoki:2021cqa}, and future experiments like HIKE~\cite{HIKE:2023ext} and HIAF~\cite{WANG:2025fmh,HIAF,Chen:2025ppt}. We propose that these experiments perform precise measurements of $\pi^-p\rightarrow K\Lambda(1405)$ and $\pi^-p\rightarrow K\Lambda(1520)$, especially the $t$-distribution cross sections at large momentum transfer (corresponding to $\theta\approx90^\circ$). Such data are essential for clarifying the production mechanisms and determining the scaling exponent $n$ with greater certainty-particularly important for $\Lambda(1405)$ as a candidate exotic hadron.

\begin{acknowledgments}
This work is supported by the National Natural Science Foundation of China under Grants No.12565018, No.12065014 and No.12247101, and by the Natural Science Foundation of Gansu Province under Grant No.22JR5RA266. We acknowledge the West Light Foundation of The Chinese Academy of Sciences, Grant No.21JR7RA201 and the ``Innovation Star" Program for Postgraduates of the Gansu Provincial Department of Education No.2025CXZX-585. X.L. is also supported by the National Natural Science Foundation of China under Grant No. 12335001, the ‘111 Center’ under Grant No. B20063, the Natural Science Foundation of Gansu Province (No. 22JR5RA389, No. 25JRRA799), the fundamental Research Funds for the Central Universities, the project for top-notch innovative talents of Gansu Province, and Lanzhou City High-Level Talent Funding.
\end{acknowledgments}

\end{document}